\documentclass[aps,prc,twocolumn,superscriptaddress]{revtex4}

\usepackage{graphicx}
\usepackage{draftcopy}
\usepackage{afterpage}

\newcommand{\Eq}[1]   {Eq.~(\ref{#1})}

\newcommand{\Fi}[1]   {Fig.~\ref{#1}}
\newcommand{\Fis}[2]  {Figs.~\ref{#1} and~\ref{#2}}
\newcommand{\Fiss}[3] {Figs.~\ref{#1},~\ref{#2}, and~\ref{#3}}
\newcommand{\Ta}[1]   {Table~\ref{#1}}

\newcommand{\Tass}[3] {Tables~\ref{#1},~\ref{#2}, and~\ref{#3}}

\newcommand{\agev}    {\mbox{$A$~GeV}}               
\newcommand{\gevc}    {\mbox{GeV$/c$}}
\newcommand{\gevcc}   {\mbox{GeV$/c^2$}}

\newcommand{\mevcc}   {\mbox{MeV$/c^2$}}
\newcommand{\rb}[1]   {\mbox{\textrm{\scriptsize #1}}}

\newcommand{\rsbox}[1]{\raisebox{1.5ex}[-1.5ex]{#1}}

\newcommand{\vzero}   {\ensuremath{\textrm{V}^{0}}}
\newcommand{\lam}     {\ensuremath{\Lambda}}
\newcommand{\lab}     {\ensuremath{\bar{\Lambda}}}  
\newcommand{\xim}     {\ensuremath{\Xi^{-}}}
\newcommand{\xizero}  {\ensuremath{\Xi^{0}}}
\newcommand{\xip}     {\ensuremath{\bar{\Xi}^{+}}}
\newcommand{\xib}     {\ensuremath{\bar{\Xi}^{0}}}
\newcommand{\sig}     {\ensuremath{\Sigma^{0}}}
\newcommand{\sib}     {\ensuremath{\bar{\Sigma}^{0}}}
\newcommand{\pimin}   {\ensuremath{\pi^-}}
\newcommand{\piplus}  {\ensuremath{\pi^+}}

\newcommand{\pt}      {\ensuremath{p_{\rb{t}}}}
\newcommand{\mt}      {\ensuremath{m_{\rb{t}}}}
\newcommand{\mtmzero} {\ensuremath{m_{\rb{t}} - m_{\rb{0}}}}
\newcommand{\mtavg}   {\ensuremath{\langle m_{\rb{t}} \rangle - m_{\rb{0}}}}

\newcommand{\dedx}    {\ensuremath{\textrm{d}E/\textrm{d}x}}
\newcommand{\dndy}    {\ensuremath{\textrm{d}N/\textrm{d}y}}

\newcommand{\nwound}  {\ensuremath{\langle N_{\rb{w}} \rangle}}
\newcommand{\navg}    {\ensuremath{\langle N \rangle}}

\newcommand{\der}     {\ensuremath{\textrm{d}}}

\newcommand{\ebeam}   {\ensuremath{E_{\rb{beam}}}}

\begin{document}



\title{System-size dependence of $\Lambda$ and $\Xi$ production in
nucleus--nucleus collisions at 40$A$ and 158\agev\ measured at the
CERN Super Proton Synchrotron}





%
%

\affiliation{NIKHEF, 
             Amsterdam, Netherlands.}
\affiliation{Department of Physics, University of Athens, 
             Athens, Greece.}
\affiliation{Comenius University, 
             Bratislava, Slovakia.}
\affiliation{KFKI Research Institute for Particle and Nuclear Physics,
             Budapest, Hungary.} 
\affiliation{MIT, 
             Cambridge, USA.}
\affiliation{Henryk Niewodniczanski Institute of Nuclear Physics, 
             Polish Academy of Sciences, 
             Cracow, Poland.}
\affiliation{Gesellschaft f\"{u}r Schwerionenforschung (GSI),
             Darmstadt, Germany.} 
\affiliation{Joint Institute for Nuclear Research, 
             Dubna, Russia.}
\affiliation{Fachbereich Physik der Universit\"{a}t, 
             Frankfurt, Germany.}
\affiliation{CERN, 
             Geneva, Switzerland.}
\affiliation{Institute of Physics \'{S}wi\c{e}tokrzyska Academy, 
             Kielce, Poland.}
\affiliation{Fachbereich Physik der Universit\"{a}t, 
             Marburg, Germany.}
\affiliation{Max-Planck-Institut f\"{u}r Physik, 
             Munich, Germany.}
\affiliation{Charles University, Faculty of Mathematics and Physics,
             Institute of Particle and Nuclear Physics, 
             Prague, Czech Republic.} 
\affiliation{Department of Physics, Pusan National University, 
             Pusan, Republic of Korea.} 
\affiliation{Nuclear Physics Laboratory, University of Washington,
             Seattle, WA, USA.} 
\affiliation{Atomic Physics Department, Sofia University St.~Kliment Ohridski, 
             Sofia, Bulgaria.} 
\affiliation{Institute for Nuclear Research and Nuclear Energy, 
             Sofia, Bulgaria.}
\affiliation{Department of Chemistry, Stony Brook Univ. (SUNYSB), 
             Stony Brook, USA.}
\affiliation{Institute for Nuclear Studies, 
             Warsaw, Poland.}
\affiliation{Institute for Experimental Physics, University of Warsaw,
             Warsaw, Poland.} 
\affiliation{Faculty of Physics, Warsaw University of Technology, 
             Warsaw, Poland.}
\affiliation{Rudjer Boskovic Institute, 
             Zagreb, Croatia.}


%
%

\author{T.~Anticic} 
\affiliation{Rudjer Boskovic Institute, 
             Zagreb, Croatia.}
\author{B.~Baatar}
\affiliation{Joint Institute for Nuclear Research, 
             Dubna, Russia.}
\author{D.~Barna}
\affiliation{KFKI Research Institute for Particle and Nuclear Physics,
             Budapest, Hungary.} 
\author{J.~Bartke}
\affiliation{Henryk Niewodniczanski Institute of Nuclear Physics, 
             Polish Academy of Sciences, 
             Cracow, Poland.}
\author{H.~Beck}
\affiliation{Fachbereich Physik der Universit\"{a}t, 
             Frankfurt, Germany.}
\author{L.~Betev}
\affiliation{CERN, 
             Geneva, Switzerland.}
\author{H.~Bia{\l}\-kowska} 
\affiliation{Institute for Nuclear Studies, 
             Warsaw, Poland.}
\author{C.~Blume}
\affiliation{Fachbereich Physik der Universit\"{a}t, 
             Frankfurt, Germany.}
\author{B.~Boimska}
\affiliation{Institute for Nuclear Studies, 
             Warsaw, Poland.}
\author{J.~Book}
\affiliation{Fachbereich Physik der Universit\"{a}t, 
             Frankfurt, Germany.}
\author{M.~Botje}
\affiliation{NIKHEF, 
             Amsterdam, Netherlands.}
\author{J.~Bracinik}
\affiliation{Comenius University, 
             Bratislava, Slovakia.}
\author{P.~Bun\v{c}i\'{c}}
\affiliation{CERN, 
             Geneva, Switzerland.}
\author{V.~Cerny}
\affiliation{Comenius University, 
             Bratislava, Slovakia.}
\author{P.~Christakoglou}
\affiliation{NIKHEF, 
             Amsterdam, Netherlands.}
\author{P.~Chung}
\affiliation{Department of Chemistry, Stony Brook Univ. (SUNYSB), 
             Stony Brook, USA.}
\author{O.~Chvala}
\affiliation{Charles University, Faculty of Mathematics and Physics,
             Institute of Particle and Nuclear Physics, 
             Prague, Czech Republic.} 
\author{J.G.~Cramer}
\affiliation{Nuclear Physics Laboratory, University of Washington,
             Seattle, WA, USA.} 
\author{P.~Csat\'{o}} 
\affiliation{KFKI Research Institute for Particle and Nuclear Physics,
             Budapest, Hungary.}
\author{P.~Dinkelaker}
\affiliation{Fachbereich Physik der Universit\"{a}t, 
             Frankfurt, Germany.}
\author{V.~Eckardt}
\affiliation{Max-Planck-Institut f\"{u}r Physik, 
             Munich, Germany.}
\author{Z.~Fodor}
\affiliation{KFKI Research Institute for Particle and Nuclear Physics,
             Budapest, Hungary.} 
\author{P.~Foka}
\affiliation{Gesellschaft f\"{u}r Schwerionenforschung (GSI),
             Darmstadt, Germany.} 
\author{V.~Friese}
\affiliation{Gesellschaft f\"{u}r Schwerionenforschung (GSI),
             Darmstadt, Germany.} 
\author{J.~G\'{a}l}
\affiliation{KFKI Research Institute for Particle and Nuclear Physics,
             Budapest, Hungary.} 
\author{M.~Ga\'zdzicki}
\affiliation{Fachbereich Physik der Universit\"{a}t, 
             Frankfurt, Germany.}
\affiliation{Institute of Physics \'{S}wi\c{e}tokrzyska Academy, 
             Kielce, Poland.}
\author{V.~Genchev}
\affiliation{Institute for Nuclear Research and Nuclear Energy, 
             Sofia, Bulgaria.}
\author{K.~Grebieszkow}
\affiliation{Faculty of Physics, Warsaw University of Technology, 
             Warsaw, Poland.}
\author{S.~Hegyi}
\affiliation{KFKI Research Institute for Particle and Nuclear Physics,
             Budapest, Hungary.} 
\author{C.~H\"{o}hne}
\affiliation{Gesellschaft f\"{u}r Schwerionenforschung (GSI),
             Darmstadt, Germany.} 
\author{K.~Kadija}
\affiliation{Rudjer Boskovic Institute, 
             Zagreb, Croatia.}
\author{A.~Karev}
\affiliation{Max-Planck-Institut f\"{u}r Physik, 
             Munich, Germany.}
\author{D.~Kresan}
\affiliation{Gesellschaft f\"{u}r Schwerionenforschung (GSI),
             Darmstadt, Germany.} 
\author{V.I.~Kolesnikov}
\affiliation{Joint Institute for Nuclear Research, 
             Dubna, Russia.}
\author{M.~Kowalski}
\affiliation{Henryk Niewodniczanski Institute of Nuclear Physics, 
             Polish Academy of Sciences, 
             Cracow, Poland.}
\author{I.~Kraus}
\affiliation{Gesellschaft f\"{u}r Schwerionenforschung (GSI),
             Darmstadt, Germany.} 
\author{M.~Kreps}
\affiliation{Comenius University, 
             Bratislava, Slovakia.}
\author{A.~Laszlo}
\affiliation{KFKI Research Institute for Particle and Nuclear Physics,
             Budapest, Hungary.} 
\author{R.~Lacey}
\affiliation{Department of Chemistry, Stony Brook Univ. (SUNYSB), 
             Stony Brook, USA.}
\author{M.~van~Leeuwen}
\affiliation{NIKHEF, 
             Amsterdam, Netherlands.}
\author{P.~L\'{e}vai}
\affiliation{KFKI Research Institute for Particle and Nuclear Physics,
             Budapest, Hungary.} 
\author{L.~Litov}
\affiliation{Atomic Physics Department, Sofia University St.~Kliment Ohridski, 
             Sofia, Bulgaria.} 
\author{B.~Lungwitz}
\affiliation{Fachbereich Physik der Universit\"{a}t, 
             Frankfurt, Germany.}
\author{M.~Makariev}
\affiliation{Institute for Nuclear Research and Nuclear Energy, 
             Sofia, Bulgaria.}
\author{A.I.~Malakhov}
\affiliation{Joint Institute for Nuclear Research, 
             Dubna, Russia.}
\author{M.~Mateev}
\affiliation{Atomic Physics Department, Sofia University St.~Kliment Ohridski, 
             Sofia, Bulgaria.} 
\author{G.L.~Melkumov}
\affiliation{Joint Institute for Nuclear Research, 
             Dubna, Russia.}
\author{C.~Meurer}
\affiliation{Fachbereich Physik der Universit\"{a}t, 
             Frankfurt, Germany.}
\author{A.~Mischke}
\affiliation{NIKHEF, 
             Amsterdam, Netherlands.}
\author{M.~Mitrovski}
\affiliation{Fachbereich Physik der Universit\"{a}t, 
             Frankfurt, Germany.}
\author{J.~Moln\'{a}r}
\affiliation{KFKI Research Institute for Particle and Nuclear Physics,
             Budapest, Hungary.} 
\author{St.~Mr\'owczy\'nski}
\affiliation{Institute of Physics \'{S}wi\c{e}tokrzyska Academy, 
             Kielce, Poland.}
\author{V.~Nicolic}
\affiliation{Rudjer Boskovic Institute, 
             Zagreb, Croatia.}
\author{G.~P\'{a}lla}
\affiliation{KFKI Research Institute for Particle and Nuclear Physics,
             Budapest, Hungary.} 
\author{A.D.~Panagiotou}
\affiliation{Department of Physics, University of Athens, 
             Athens, Greece.}
\author{D.~Panayotov}
\affiliation{Atomic Physics Department, Sofia University St.~Kliment Ohridski, 
             Sofia, Bulgaria.} 
\author{A.~Petridis}
\altaffiliation{deceased}
\affiliation{Department of Physics, University of Athens, 
             Athens, Greece.}
\author{W.~Peryt}
\affiliation{Faculty of Physics, Warsaw University of Technology, 
             Warsaw, Poland.}
\author{M.~Pikna}
\affiliation{Comenius University, 
             Bratislava, Slovakia.}
\author{J.~Pluta}
\affiliation{Faculty of Physics, Warsaw University of Technology, 
             Warsaw, Poland.}
\author{D.~Prindle}
\affiliation{Nuclear Physics Laboratory, University of Washington,
             Seattle, WA, USA.} 
\author{F.~P\"{u}hlhofer}
\affiliation{Fachbereich Physik der Universit\"{a}t, 
             Marburg, Germany.}
\author{R.~Renfordt}
\affiliation{Fachbereich Physik der Universit\"{a}t, 
             Frankfurt, Germany.}
\author{C.~Roland}
\affiliation{MIT, 
             Cambridge, USA.}
\author{G.~Roland}
\affiliation{MIT, 
             Cambridge, USA.}
\author{M.~Rybczy\'nski}
\affiliation{Institute of Physics \'{S}wi\c{e}tokrzyska Academy, 
             Kielce, Poland.}
\author{A.~Rybicki}
\affiliation{Henryk Niewodniczanski Institute of Nuclear Physics, 
             Polish Academy of Sciences, 
             Cracow, Poland.}
\author{A.~Sandoval}
\affiliation{Gesellschaft f\"{u}r Schwerionenforschung (GSI),
             Darmstadt, Germany.} 
\author{N.~Schmitz}
\affiliation{Max-Planck-Institut f\"{u}r Physik, 
             Munich, Germany.}
\author{T.~Schuster}
\affiliation{Fachbereich Physik der Universit\"{a}t, 
             Frankfurt, Germany.}
\author{P.~Seyboth}
\affiliation{Max-Planck-Institut f\"{u}r Physik, 
             Munich, Germany.}
\author{F.~Sikl\'{e}r}
\affiliation{KFKI Research Institute for Particle and Nuclear Physics,
             Budapest, Hungary.} 
\author{B.~Sitar}
\affiliation{Comenius University, 
             Bratislava, Slovakia.}
\author{E.~Skrzypczak}
\affiliation{Institute for Experimental Physics, University of Warsaw,
             Warsaw, Poland.} 
\author{M.~Slodkowski}
\affiliation{Faculty of Physics, Warsaw University of Technology, 
             Warsaw, Poland.}
\author{G.~Stefanek}
\affiliation{Institute of Physics \'{S}wi\c{e}tokrzyska Academy, 
             Kielce, Poland.}
\author{R.~Stock}
\affiliation{Fachbereich Physik der Universit\"{a}t, 
             Frankfurt, Germany.}
\author{C.~Strabel}
\affiliation{Fachbereich Physik der Universit\"{a}t, 
             Frankfurt, Germany.}
\author{H.~Str\"{o}bele}
\affiliation{Fachbereich Physik der Universit\"{a}t, 
             Frankfurt, Germany.}
\author{T.~Susa}
\affiliation{Rudjer Boskovic Institute, 
             Zagreb, Croatia.}
\author{I.~Szentp\'{e}tery}
\affiliation{KFKI Research Institute for Particle and Nuclear Physics,
             Budapest, Hungary.} 
\author{J.~Sziklai}
\affiliation{KFKI Research Institute for Particle and Nuclear Physics,
             Budapest, Hungary.} 
\author{M.~Szuba}
\affiliation{Faculty of Physics, Warsaw University of Technology, 
             Warsaw, Poland.}
\author{P.~Szymanski}
\affiliation{Institute for Nuclear Studies, 
             Warsaw, Poland.}
\author{M.~Utvi\'{c}}
\affiliation{Fachbereich Physik der Universit\"{a}t, 
             Frankfurt, Germany.}
\author{D.~Varga}
\affiliation{KFKI Research Institute for Particle and Nuclear Physics,
             Budapest, Hungary.} 
\affiliation{CERN, 
             Geneva, Switzerland.}
\author{M.~Vassiliou}
\affiliation{Department of Physics, University of Athens, 
             Athens, Greece.}
\author{G.I.~Veres}
\affiliation{KFKI Research Institute for Particle and Nuclear Physics,
             Budapest, Hungary.} 
\affiliation{MIT, 
             Cambridge, USA.}
\author{G.~Vesztergombi}
\affiliation{KFKI Research Institute for Particle and Nuclear Physics,
             Budapest, Hungary.}
\author{D.~Vrani\'{c}}
\affiliation{Gesellschaft f\"{u}r Schwerionenforschung (GSI),
             Darmstadt, Germany.} 
\author{Z.~W{\l}odarczyk}
\affiliation{Institute of Physics \'{S}wi\c{e}tokrzyska Academy, 
             Kielce, Poland.}
\author{A.~Wojtaszek}
\affiliation{Institute of Physics \'{S}wi\c{e}tokrzyska Academy, 
             Kielce, Poland.}
\author{I.K.~Yoo}
\affiliation{Department of Physics, Pusan National University, 
             Pusan, Republic of Korea.} 



\collaboration{The NA49 Collaboration}
\noaffiliation


\begin{abstract}

Results on \lam, \lab, and \xim\ production in centrality selected Pb+Pb reactions
at 40$A$ and 158\agev\ and in near-central C+C and Si+Si reactions at 158\agev\ 
are presented.  Transverse mass spectra, rapidity spectra, and multiplicities in
dependence of the system size are discussed.  Comparisons to transport models
(UrQMD2.3 and HSD) and to the core-corona approach are performed.  While \lam\ 
and \lab\ production can be described by transport models reasonably well, these 
models do not agree with the \xim\ measurements.  On the other hand, the 
core-corona picture fits very well the system-size dependence of \lam\ and \xim, 
while it agrees less well with the \lab\ data.  
\end{abstract}




\maketitle

\section{Introduction}

The production of strange particles has always been a key observable in 
heavy-ion reactions and its enhancement was one of the first suggested 
signatures for quark-gluon plasma (QGP) formation \cite{RAFELSKI}.  The 
predicted enhancement of strangeness production in nucleus--nucleus 
collisions relative to proton--proton reactions was established 
experimentally some time ago \cite{NA35LAM,NA35STR}.  It was also found 
that this enhancement is increasing with the strangeness content of the 
particle type \cite{WA97HYP,NA57HY158}.

However, there are several aspects that make a straightforward 
interpretation of the experimental results difficult.  One of them is the 
fact, that the strangeness enhancement seems to increase towards lower 
energies \cite{E802STR,E895HYP,NA49EDHYP}.  Another open issue and the
topic of this publication is the dependence of the enhancement on the
system size.  The latter is quantified by the number of ``wounded''
nucleons from the colliding nuclei that participate in the reaction.  A 
previous analysis investigated the pion, kaon, $\phi$, and \lam\ production 
in (near-)central nucleus--nucleus collisions of nuclei with different mass 
numbers \cite{NA49CCSISI}.  It turned out that the enhancement sets in 
already for very small systems and seems to saturate for nuclei with mass 
number above 30, i.~e. number of participants above 60.  The present study 
extends the investigation of the enhancement to non-central Pb+Pb collisions
and to the production of \lab\ and \xim\ hyperons.

In the context of statistical models, which generally have been quite 
successful in describing particles yields, the experimentally observed
strangeness enhancement for large systems is due to the increase of the 
reaction volume, which weakens the influence of strangeness conservation
on the production rate \cite{DANOS}.  In \cite{TOUNSI} this has been 
modeled by the transition from a canonical ensemble to a grand-canonical 
one.  For comparison of the corresponding statistical model results to 
experimental data, the relation between the size of the ensemble volume 
$V$ and the experimentally accessible number of wounded nucleons \nwound\ 
has been assumed to be $V = (V_{0} / 2) \: \nwound$.  The parameter 
$V_{0}$, which accounts for the locality of the strangeness conservation, 
is usually fixed to $\approx$~7~fm$^{3}$ \cite{TOUNSI}.  However, this 
approach does not provide a satisfactory description of the data 
\cite{NA49CCSISI}.  A better agreement with global strangeness production 
at the SPS can be achieved by deriving the ensemble volume from a 
percolation of elementary clusters \cite{CLAUDIA}.  A similar, although 
simplified, line of argument is put forward in the so-called core-corona 
picture \cite{WERNER1}.  Here a heavy-ion collision is considered a 
superposition of a hot and dense core and a low density peripheral corona 
region.  While the core region corresponds to a large fireball, which 
experiences collective expansion and for which particle production should 
be describable via a large volume canonical ensemble, or equivalently by
a grand-canonical statistical ensemble, the corona is supposed to exhibit 
the features of simple nucleon-nucleon collisions.  The contributions of 
the core and the corona depend on both the size of the colliding nuclei 
and the centrality of the collision and can be determined via Glauber 
model calculations \cite{GLAUBER}.  This concept has recently been applied 
quite successfully to the system-size dependence of strangeness production 
at RHIC energies \cite{BECATTINI1,BECATTINI3,TIMMINS,WERNER2}.

The comparison of measurements with hadron--string transport models, such
as UrQMD or HSD, provides insight into the question whether nucleus--nucleus 
collisions can be described in a hadronic scenario or whether a contribution 
from an additional partonic phase is needed.  Even though these models are 
not able to describe the enhancement of multi-strange particle yields in 
central nucleus--nucleus collisions \cite{NA49EDHYP}, their predictions are 
generally close to the measured values for particles containing a single 
strange quark.  Their comparison to the measured system-size dependence 
might therefore reveal effects that go beyond the dominating influence of 
the reaction geometry, such as multi-pion fusion processes and, in the case 
of anti-baryons, absorption in dense nuclear matter.


\section{Data analysis}

%
\begin{figure}[t]
\begin{center}
\includegraphics[width=\linewidth]{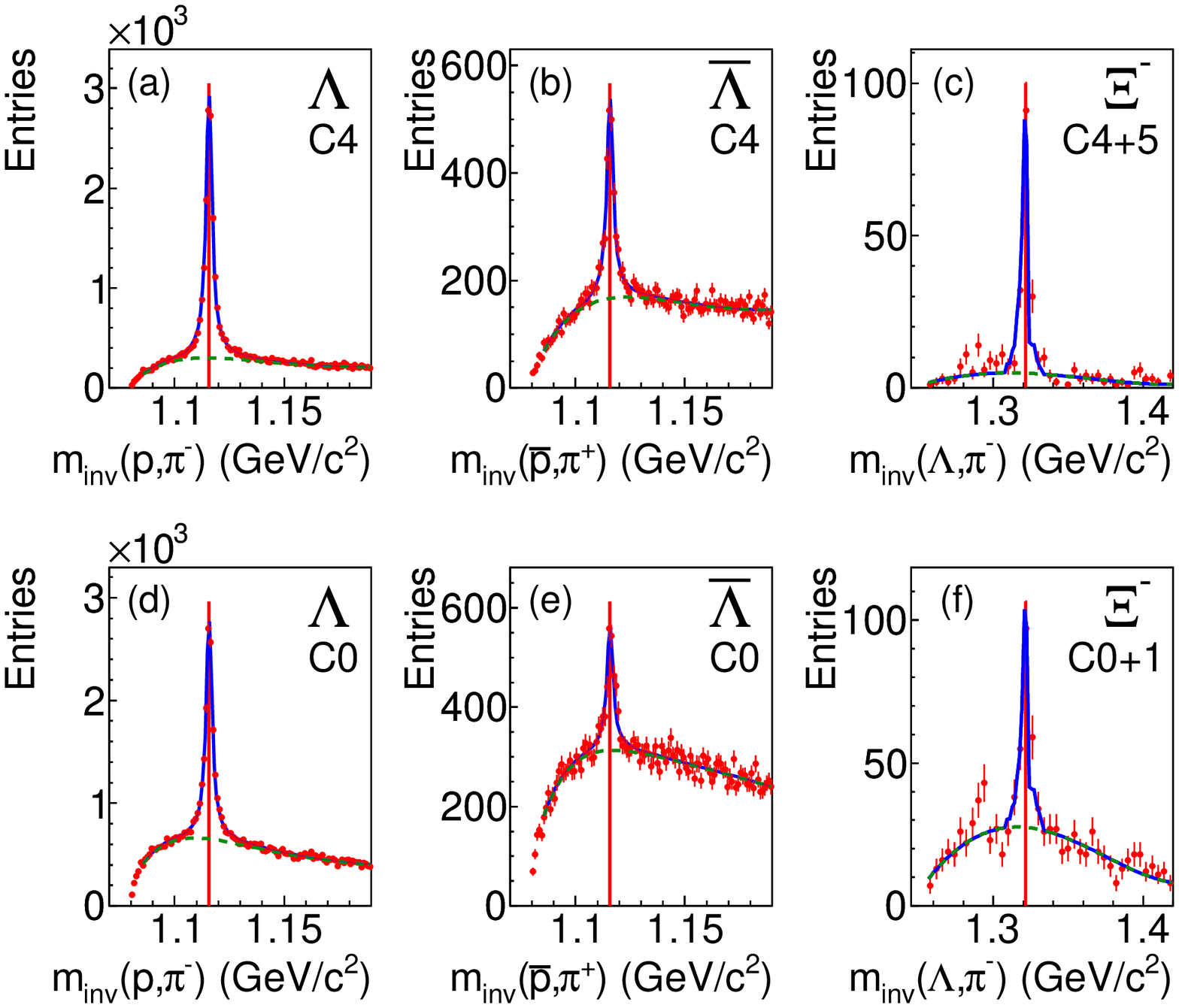}
\end{center}
\caption{(color online) The invariant mass distributions of all measured 
\lam, \lab, and \xim\ candidates in two centrality classes of Pb+Pb 
collisions at 158\agev.  The upper row (a -- c) shows the most peripheral 
centrality class, the lower row (d -- f) the most central one.  The full 
curves represent a fit to signal and background as described in the text, 
while the dashed curves show the background only.  The vertical lines 
denote the literature values
of the masses \cite{PDG}.
}
\label{fig:minv_158} 
\end{figure} 
%

%
\begin{figure}[t]
\begin{center}
\begin{minipage}[b]{0.49\linewidth}
\begin{center}
\includegraphics[width=\linewidth]{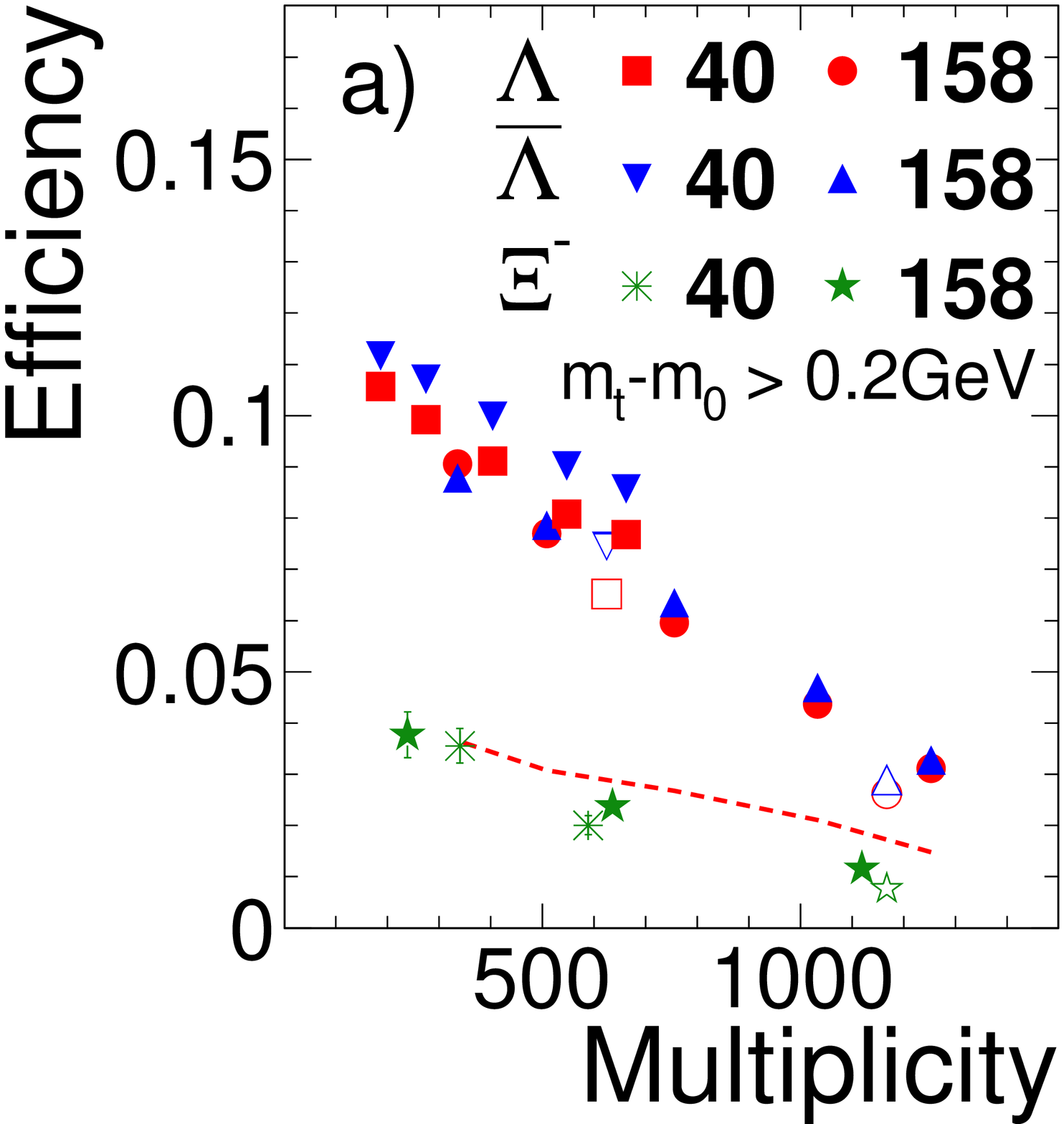}
\end{center}
\end{minipage}
\begin{minipage}[b]{0.49\linewidth}
\begin{center}
\includegraphics[width=\linewidth]{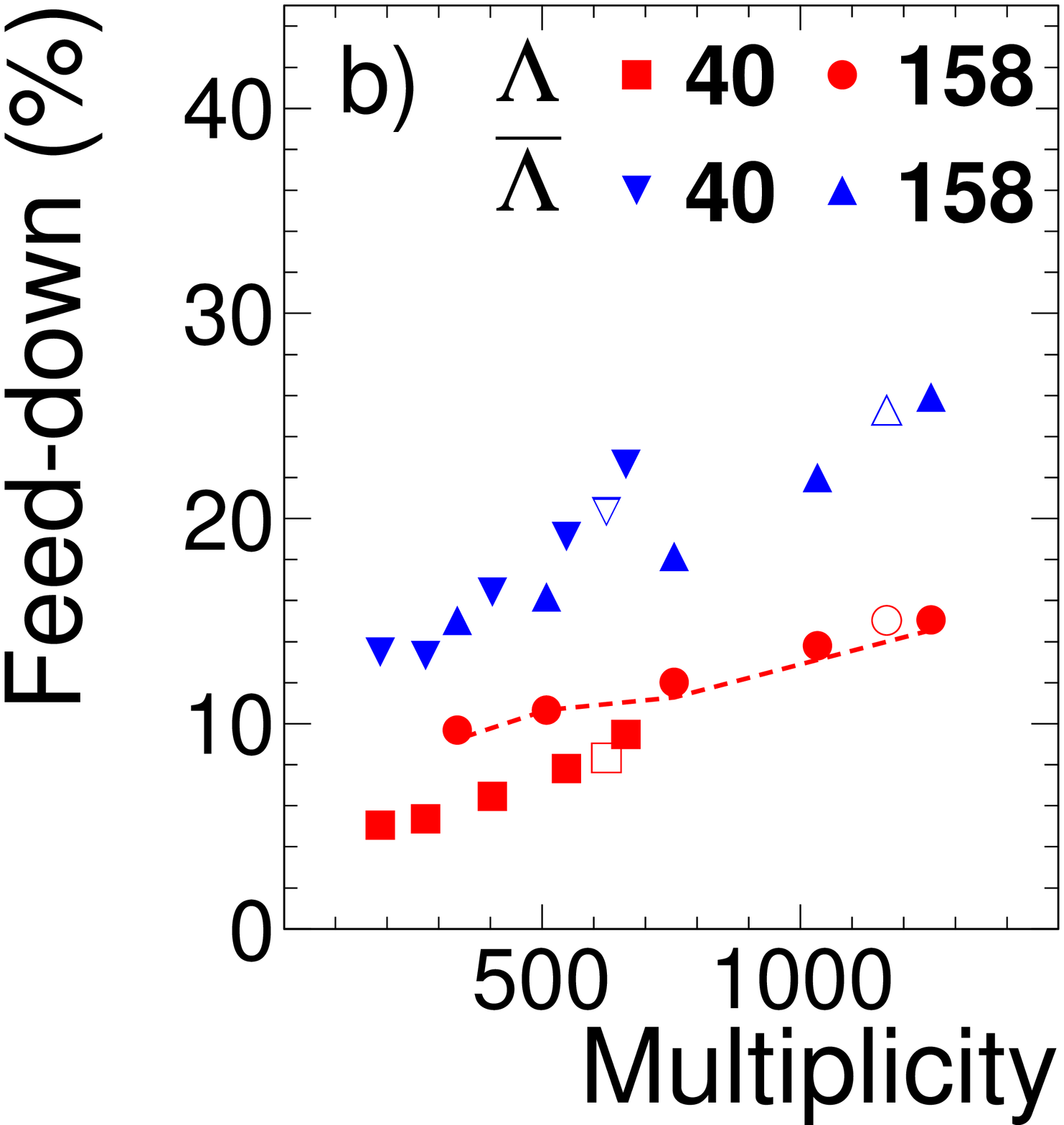}
\end{center}
\end{minipage}
\end{center}
\caption{(color online) The \pt-integrated reconstruction efficiency (a) and 
feed-down contribution (b) at mid-rapidity (\lam, \lab: $|y| < 0.4$, 
\xim: $|y| < 0.5$) as a function of centrality as characterized by the 
measured charged track multiplicity for minimum bias Pb+Pb reactions at 40$A$ 
and 158\agev.  Open symbols correspond to the values for central Pb+Pb 
collisions \cite{NA49EDHYP}.  The dashed lines represent the \lam\ efficiency 
or feed-down contribution respectively, at 158\agev, resulting from a 
different analysis strategy (see text).
}
\label{fig:eff_fd_corr} 
\end{figure} 
%

\subsection{\label{sec:exp} Experimental setup and data sets}

The data were taken with the NA49 large acceptance hadron spectrometer
at the CERN SPS.  A detailed description of the apparatus can be found 
in~\cite{NA49NIM}.  With this detector tracking is performed by four
large-volume Time Projection Chambers (TPCs) in a wide range of phase
space.  Two of these are positioned inside two superconducting dipole
magnets.  In order to ensure a similar detector acceptance in the 
center-of-mass system for all datasets, the magnetic field was reduced 
for the 40\agev\ Pb beam by a factor of 4.  A measurement of the specific 
energy loss \dedx\ in the TPC gas with a typical resolution of 4~\% 
provides particle identification at forward rapidities.  Time-of-flight 
detectors improve the particle identification at mid-rapidity.  The 
centrality of a given reaction is determined via the energy measured in 
the projectile fragmentation region by the Zero Degree Calorimeter (ZDC) 
positioned downstream of the TPCs.  A collimator in front of the ZDC 
reduces the acceptance of the calorimeter to the phase space of the 
projectile fragments and spectator nucleons. 

We present in this paper an analysis of centrality selected Pb+Pb events 
taken with a minimum bias trigger at beam energies of 40$A$ and 158\agev\ 
in the years 1999 and 2000, and of near-central C+C and Si+Si events 
measured at 158\agev\ in the year 1998.  The properties of the different 
datasets are summarized in Tables~\ref{tab:mbdatasets} and \ref{tab:frdatasets}.  
While for the Pb+Pb collisions the primary SPS beam was used, the C and Si 
ions were created by fragmenting the original Pb beam.  By tuning the 
magnetic rigidity in the beam line ($Z/A = 0.5$) and analyzing the specific 
energy loss in scintillation detectors, the corresponding ion species could 
be selected.  In the following, the carbon beam is defined as $Z = 6$ ions 
and the silicon beam as a mixture of $Z = 13 - 15$.  Two carbon targets with 
thicknesses of 3~mm and 10~mm (561~mg/cm$^{2}$ and 1840~mg/cm$^{2}$, 
respectively) and a silicon target with a thickness of 5~mm (1170~mg/cm$^{2}$) 
were used.  Further details on the analysis of the C+C and Si+Si datasets 
can be found in \cite{NA49CCSISI}.  For the study of the minimum bias Pb+Pb 
interactions targets with a thickness of 200~$\mu$m (224~mg/cm$^{2}$) were 
installed.  The minimum bias trigger is defined by a gas Cherenkov counter 
that vetoes non-interacting projectiles.  Centrality classes C0~--~C4 are
defined by consecutive intervals of spectator energy as measured in the
ZDC calorimeter.  Simulated events from the VENUS~4.12 event generator 
\cite{VENUS} were used to relate this energy to the number of wounded 
nucleons \nwound\ as given by the implemented Glauber model calculation
\cite{GLAUBER}.  The background from non-target interactions is 
substantially reduced by applying cuts on the reconstructed position of 
the primary vertex.  After these cuts the centrality classes C0~--~C2 are 
free of background events, while the more peripheral classes C3 and C4 have 
a contamination of less than 2~\% and 5~\%, respectively.

%
\begin{table}[tbh]
\caption{
Summary of the analyzed Pb+Pb datasets.  The centrality is quantified 
by the fraction of the total inelastic cross section.  \nwound\ is the 
average number of wounded nucleons per event and $\sigma(\nwound)$ 
the widths of the corresponding distributions.  For the 158\agev\ 
dataset also the fraction $f(\nwound)$ of nucleons that scatter more 
than once is given \cite{WERNER2}.  $N_{\rb{Event}}$ is the number of 
accepted events.
}
\begin{ruledtabular}
\begin{tabular}{ccccccc}
  \ebeam            &
  Class             & 
  Centrality        &
  \nwound           &
  $\sigma(\nwound)$ &
  $f(\nwound)$      &
  $N_{\rb{Event}}$                                                   \\ 
  (\agev)           &
                    & 
  (\%)              &
                    &
                    &
                    &
                                                                     \\ 
                                                                     \hline
40       & C0 &  0.0 --  5.0 & 351$\pm$3 & 16$\pm$1 & ---  & 26k \\
         & C1 &  5.0 -- 12.5 & 290$\pm$4 & 21$\pm$2 & ---  & 45k \\
         & C2 & 12.5 -- 23.5 & 210$\pm$6 & 22$\pm$2 & ---  & 66k \\
         & C3 & 23.5 -- 33.5 & 142$\pm$8 & 22$\pm$3 & ---  & 62k \\
         & C4 & 33.5 -- 43.5 &  93$\pm$7 & 21$\pm$2 & ---  & 63k \\
                                                                     \hline
158      & C0 &  0.0 --  5.0 & 352$\pm$3 & 13$\pm$1 & 0.89 & 14k \\
         & C1 &  5.0 -- 12.5 & 281$\pm$4 & 18$\pm$2 & 0.85 & 23k \\
         & C2 & 12.5 -- 23.5 & 196$\pm$6 & 19$\pm$2 & 0.80 & 36k \\
         & C3 & 23.5 -- 33.5 & 128$\pm$8 & 19$\pm$3 & 0.74 & 33k \\
         & C4 & 33.5 -- 43.5 &  85$\pm$7 & 18$\pm$2 & 0.68 & 33k \\
\end{tabular}
\end{ruledtabular}
\label{tab:mbdatasets}
\end{table}
%

%
\begin{table}[tbh]
\caption{
Summary of the analyzed near-central C+C and Si+Si datasets.  The 
centrality is quantified by the fraction of the total inelastic cross 
section.  \nwound\ is the average number of wounded nucleons per 
event and $N_{\rb{Event}}$ the number of accepted events.
}
\begin{ruledtabular}
\begin{tabular}{clccc}
  \ebeam           &
  Reaction         & 
  Centrality       &
  \nwound          & 
  $N_{\rb{Event}}$                                    \\ 
  (\agev)          &
                   & 
  (\%)             &
                   & 
                                                      \\ 
                                                      \hline
158      & C+C   &  0.0 -- 15.3 & 14$\pm$2 & 250k \\
         & Si+Si &  0.0 -- 12.2 & 37$\pm$3 & 200k \\
\end{tabular}
\end{ruledtabular}
\label{tab:frdatasets}
\end{table}
%

\subsection{\label{sec:rec} \lam, \lab, and \xim\ reconstruction}

%
\begin{figure}[t]
\begin{center}
\includegraphics[width=0.9\linewidth]{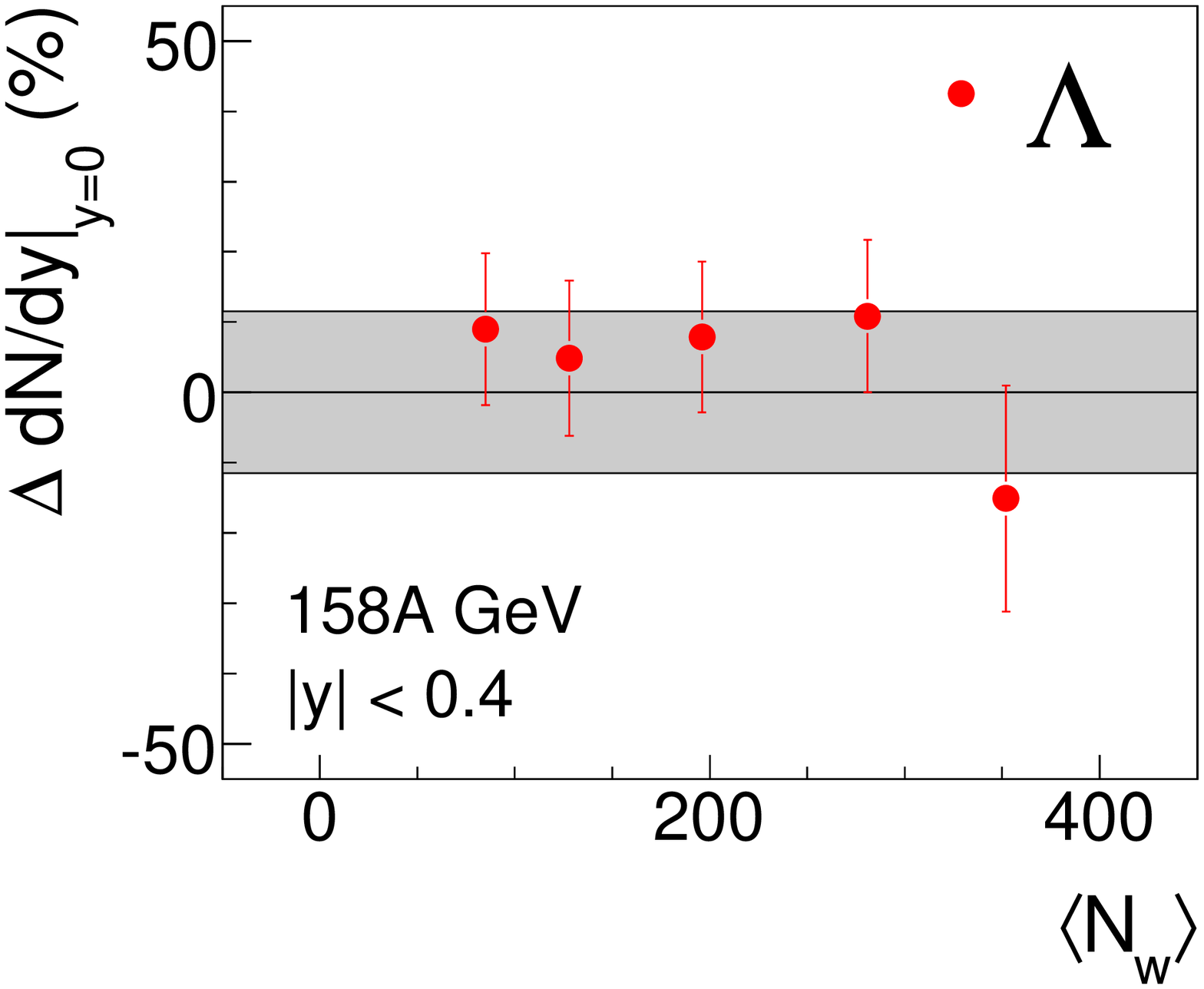}
\end{center}
\caption{(color online) The differences between the fully corrected 
\dndy\ values at mid-rapidity for \lam\ at 158\agev\ as extracted with 
the standard procedure and the alternative analysis strategy (see section 
\ref{sec:eff}) for the different centrality bins.  The gray area
illustrates the systematic error estimate.
}
\label{fig:syserr} 
\end{figure} 
%

%
\begin{figure*}[t]
\begin{center}
\begin{minipage}[b]{0.32\linewidth}
\begin{center}
\includegraphics[width=55mm]{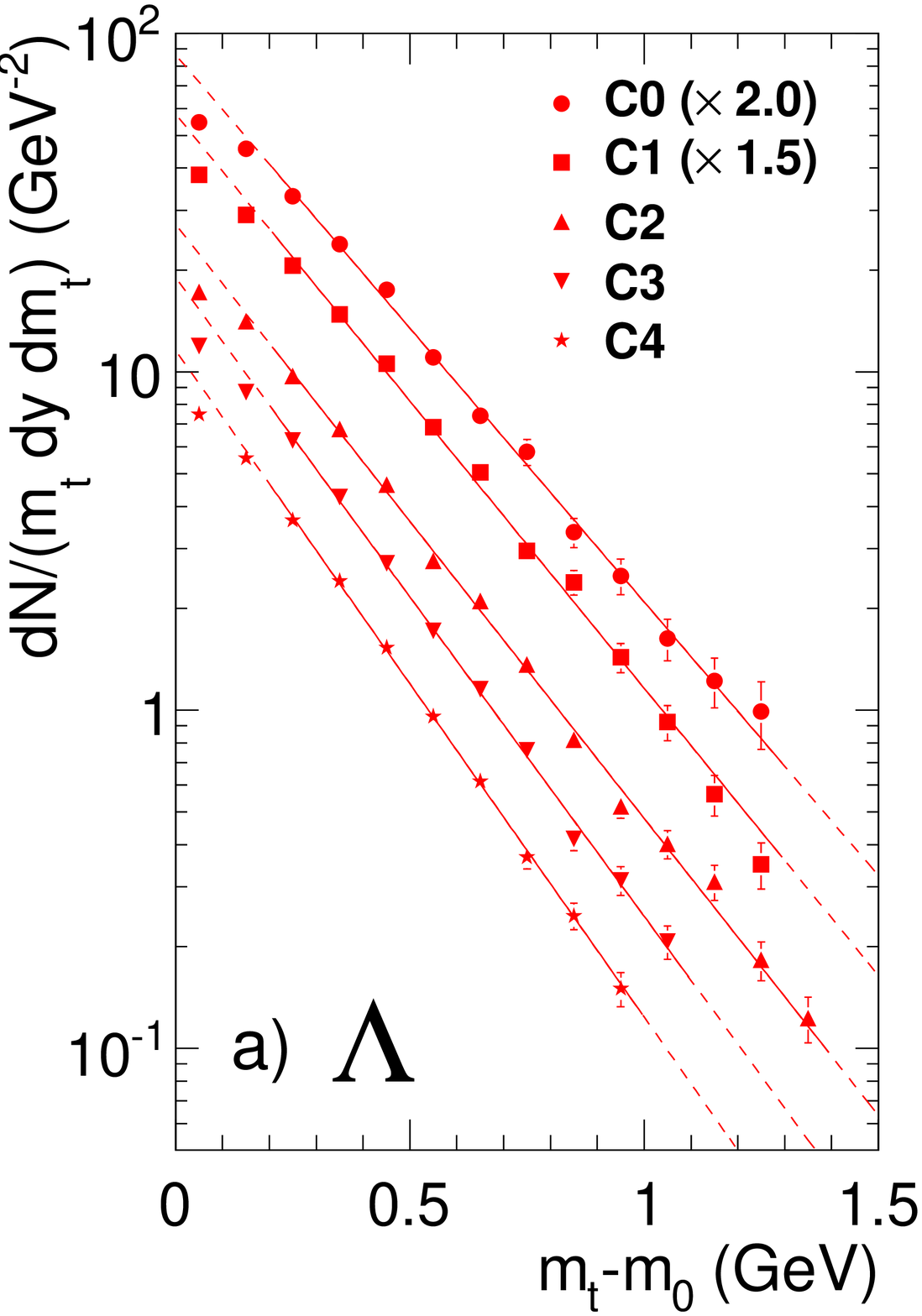}
\end{center}
\end{minipage}
\begin{minipage}[b]{0.32\linewidth}
\begin{center}
\includegraphics[width=55mm]{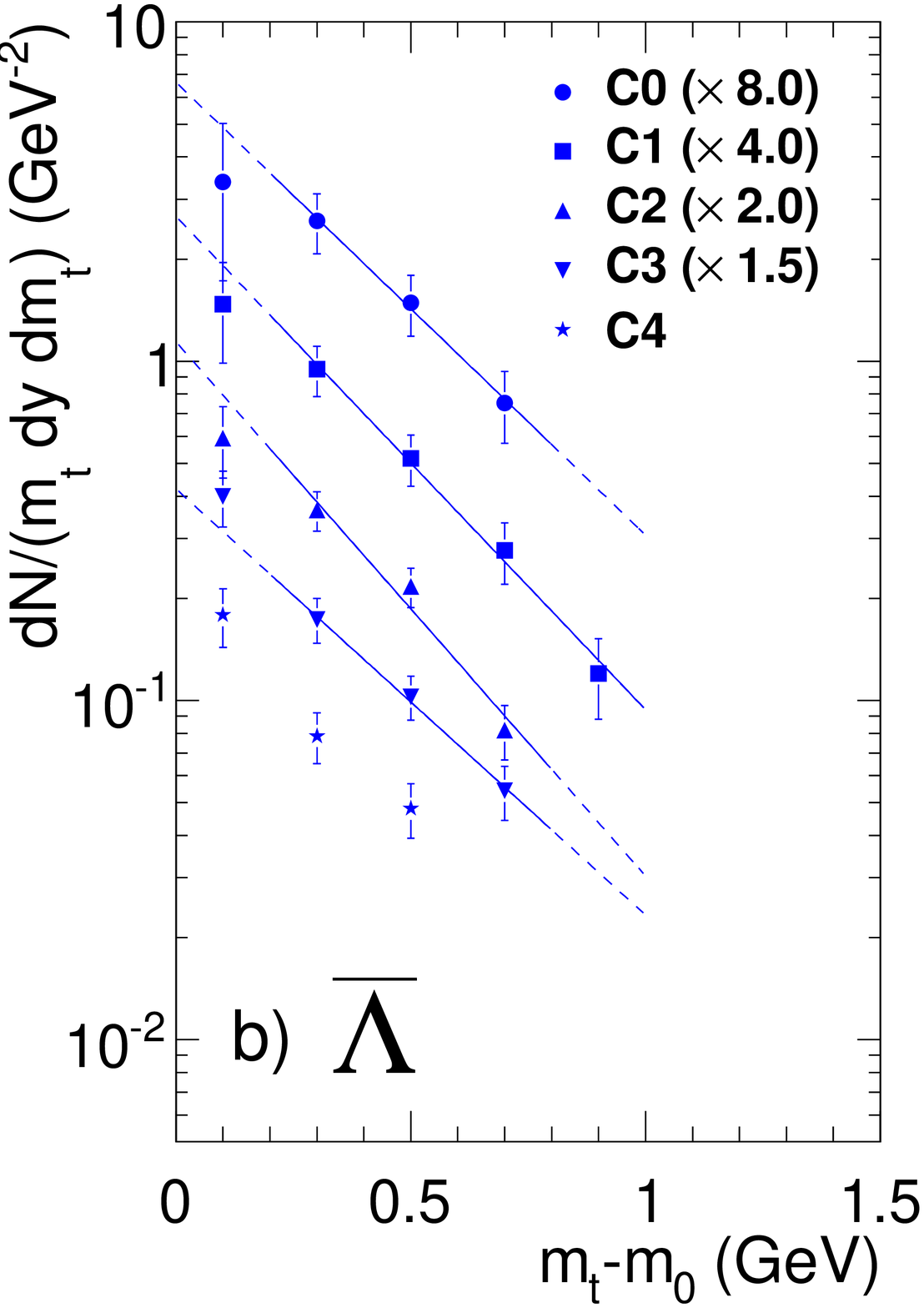}
\end{center}
\end{minipage}
\begin{minipage}[b]{0.32\linewidth}
\begin{center}
\includegraphics[width=55mm]{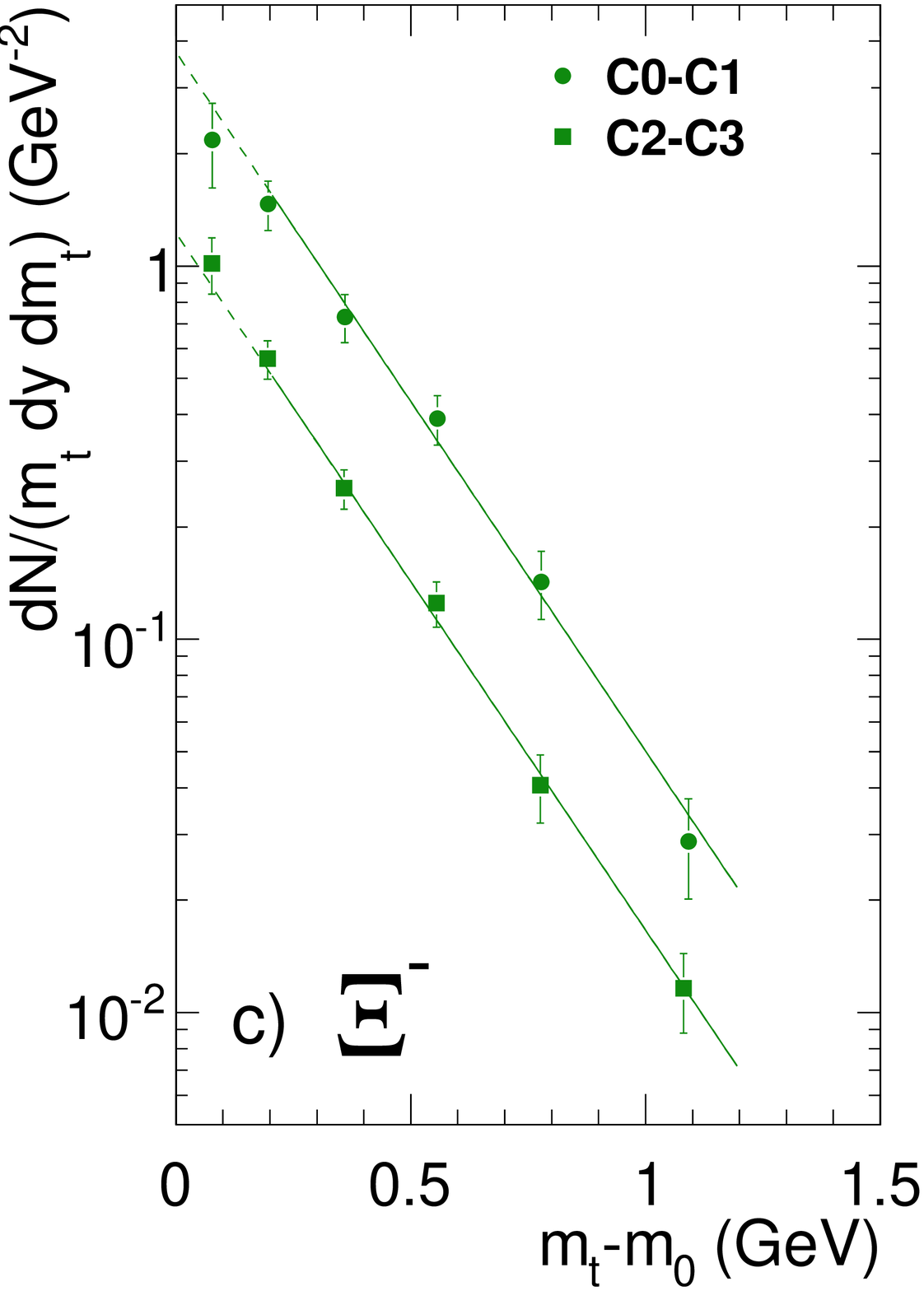}
\end{center}
\end{minipage}
\end{center}
\caption{ (color online)
The transverse mass spectra of \lam\ (a), \lab\ (b), and \xim\ (c) at 
mid-rapidity (\lam/\lab: $|y| < 0.4$, \xim: $|y| < 0.5$) for Pb+Pb 
reactions at 40\agev\ in different centrality bins.  Some of the data 
points are scaled for clarity.  Only statistical errors are shown.  
The solid/dashed lines represent fits with an exponential, where the 
solid parts denote the \mt\ ranges in which the fits were performed.
}
\label{fig:lm_lb_xi_mt_40} 
\end{figure*}
%

The reconstruction of \lam, \lab, and \xim\ follows the same procedures
as employed in a previous analysis \cite{NA49EDHYP}.  A detailed
description of the methods, together with a list of all applied cuts, 
can thus be found there.  Here we just summarize the basic principles.

\lam\ and \lab\ hyperons were reconstructed from their charged decays 
$\lam \rightarrow \pimin + \textrm{p}$ and 
$\lab \rightarrow \piplus + \bar{\textrm{p}}$ (branching ratio 63.9~\% 
\cite{PDG}).  Candidate pairs were formed by combining all reconstructed 
tracks of positively with all tracks of negatively charged particles.  
Pairs with a distance of closest approach (DCA) of less than 0.5~cm 
anywhere between the position of the first measured points on the tracks 
and the target plane are treated as possible \vzero\ candidates.  The 
(anti-)protons are identified via their specific energy loss (\dedx) in 
the TPCs which removes the background resulting from a wrong mass 
assignment.  The combinatorial background is further reduced by applying 
additional cuts to \vzero\ candidates.  These include a minimal distance 
of the reconstructed decay vertex position to the main interaction vertex 
and the requirement that the reconstructed momentum vectors of the \lam\ 
(\lab) candidates should point back to the interaction vertex.  Due to the 
lower multiplicities in C+C and Si+Si reactions, the combinatorial 
background for \lam\ and \lab\ is much smaller.  Therefore, the cuts have 
been relaxed compared to the ones in the analysis of the Pb+Pb data.  
\xim\ candidates were identified via the decay channel 
$\xim \rightarrow \lam + \pimin$ which has a branching ratio of 99.9~\% 
\cite{PDG}.  To reconstruct the \xim, \lam\ candidates were selected in an 
invariant mass window of 1.101~--~1.131~\gevcc\ and combined with all 
measured negatively charged particles in the event.  The \lam\ candidates
were subjected to the same selection criteria as used in the \lam\ analysis,
except for the momentum pointing cut.  The reconstructed \xim\ candidates 
were required to point back to the interaction vertex.  In order to further
reduce the number of fake \xim, the trajectories of the negatively charged
pions from \lam\ and \xim\ decays were intersected with the target plane
and the distances of the resulting positions to the main interaction vertex 
were required to be larger than a minimum value.

The invariant mass spectra were measured in bins of \pt, (\mtmzero), $y$, 
as well as centrality, and fitted to the sum of a polynomial and a signal 
distribution, the latter determined from simulation.  The raw yields of 
\lam, \lab, and \xim\ were obtained by subtracting the fitted background 
and integra\-ting the remaining signal distributions in a mass window of 
$\pm 11$~\mevcc\ ($\pm 10$~\mevcc) around the nominal $\Lambda$ ($\Xi$) 
mass.  Typical mass resolutions for \lam\ ($\Xi$), as obtained from a
fit with a Gaussian, are $\sigma_{\rb{m}} \approx 2$~MeV$/c^{2}$ 
(4~MeV$/c^{2}$).  Figure~\ref{fig:minv_158} shows examples of invariant 
mass distributions for two centrality classes of Pb+Pb collisions at 
158\agev\ together with the corresponding fitted curves.  

\subsection{\label{sec:eff} Correction for acceptance, reconstruction 
inefficiency and feed-down}

%
\begin{figure*}[t]
\begin{center}
\begin{minipage}[b]{0.32\linewidth}
\begin{center}
\includegraphics[width=55mm]{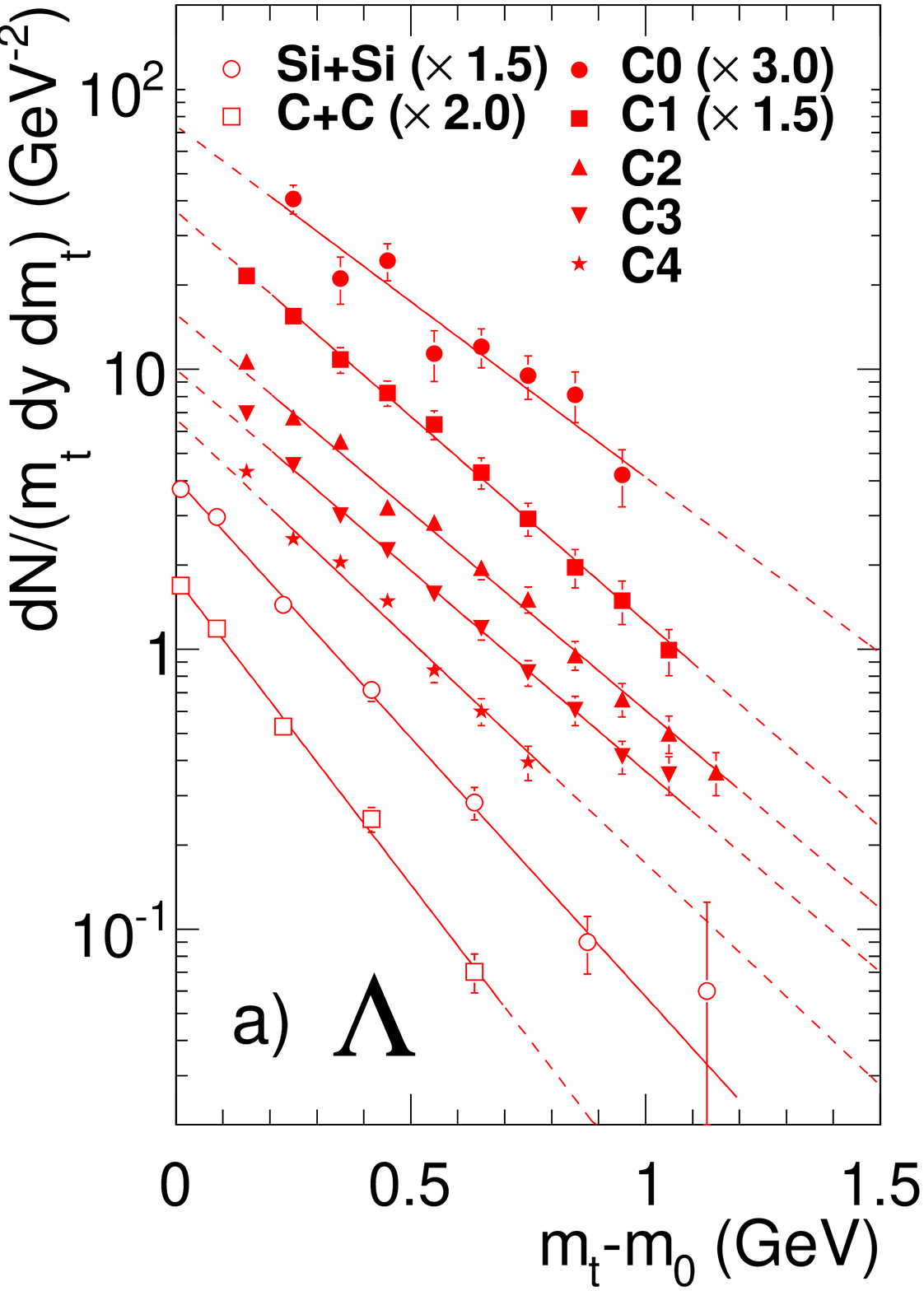}
\end{center}
\end{minipage}
\begin{minipage}[b]{0.32\linewidth}
\begin{center}
\includegraphics[width=55mm]{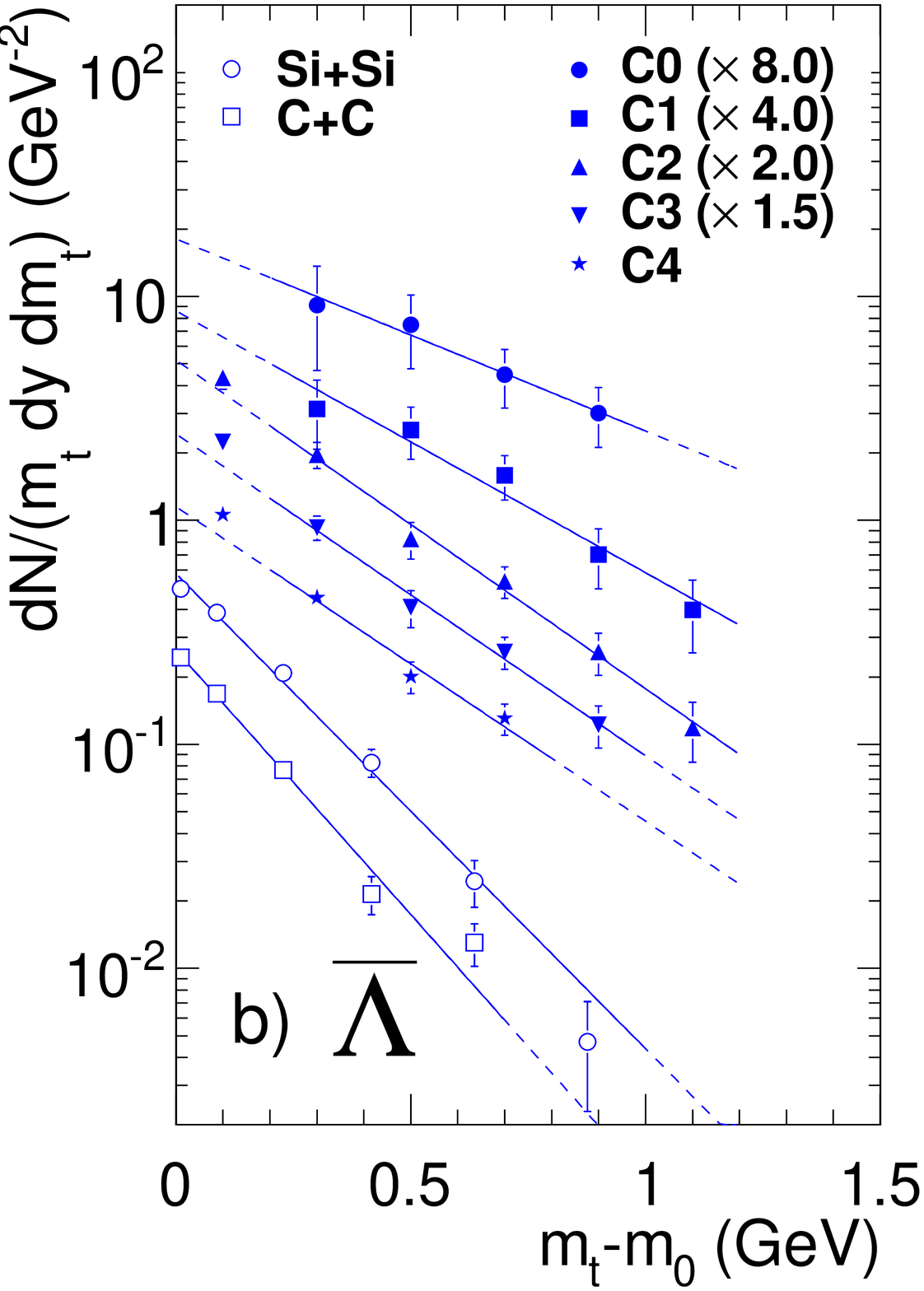}
\end{center}
\end{minipage}
\begin{minipage}[b]{0.32\linewidth}
\begin{center}
\includegraphics[width=55mm]{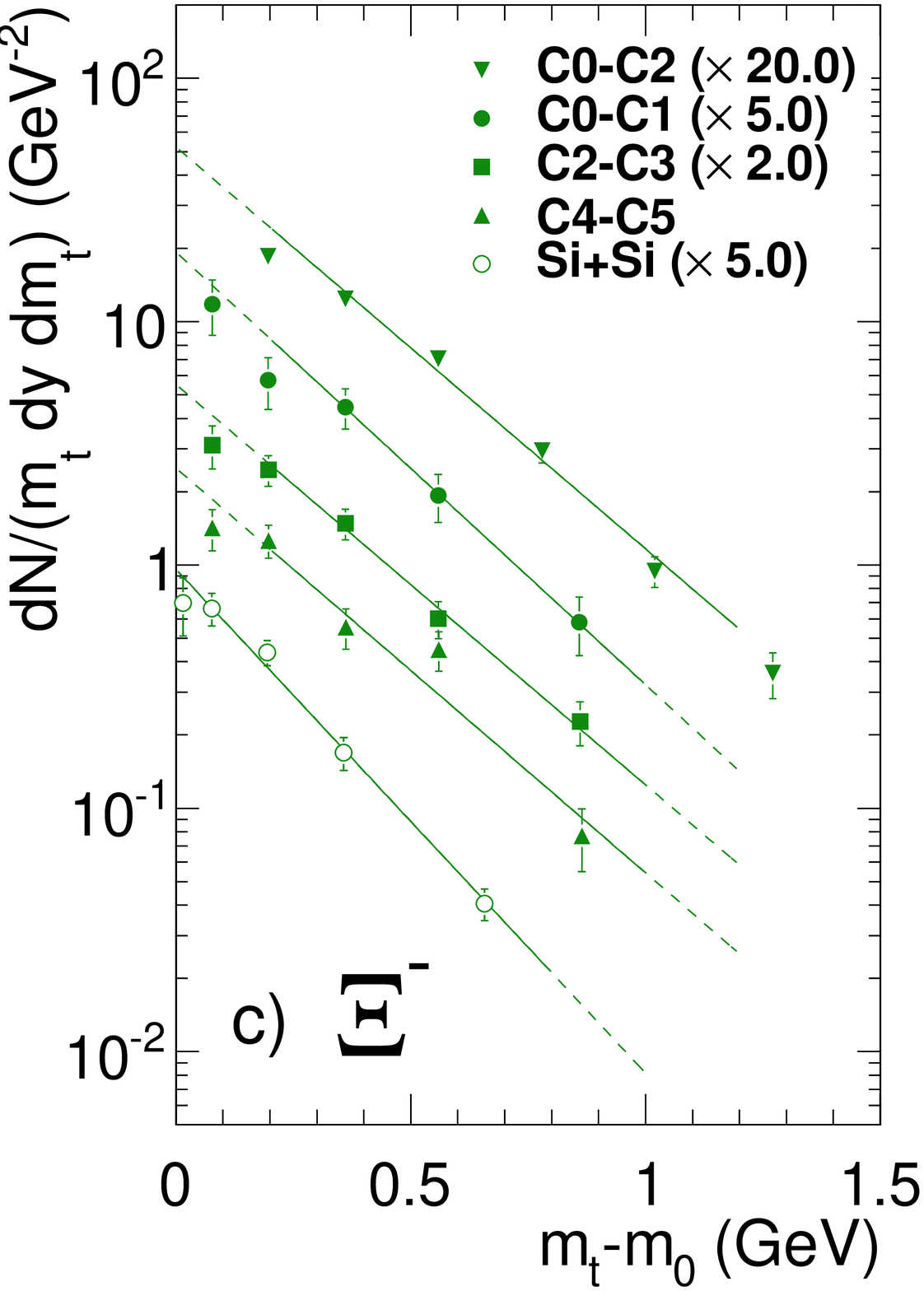}
\end{center}
\end{minipage}
\end{center}
\caption{(color online)
The transverse mass spectra of \lam\ (a), \lab\ (b), and \xim\ (c) at 
mid-rapidity (\lam/\lab: $|y| < 0.4$, \xim: $|y| < 0.5$) for Pb+Pb 
reactions at 158\agev\ in different centrality bins, and in 
near-central C+C and Si+Si collisions at 158\agev.  Some of the data 
points are scaled for clarity.  Only statistical errors are shown.  
The solid/dashed lines represent fits with an exponential, where the 
solid parts denote the \mt\ ranges in which the fits were performed.
The \lam\ spectra for C+C and Si+Si are taken from \cite{NA49CCSISI}.}
\label{fig:lm_lb_xi_mt_158} 
\end{figure*}
%

Detailed simulations were made to correct the yields for geometrical
acceptance and inefficiencies of the reconstruction procedure.  As input
to these simulations samples of \lam, \lab\ and \xim\ were generated 
with \mt~distributions according to:
\begin{equation}
  \label{eq:expo}
  \!\! 
  \frac{1}{\mt} \frac{\der N}{\der \mt} 
  \propto
  \exp \left( - \frac{\mt}{T} \right) .
\end{equation}
Here $\pt$ is the transverse momentum, $\mt = \sqrt{\pt^{2} + m^{2}}$,
and $T$ the inverse slope parameter.  In case of the Pb+Pb collisions
the $T$ parameter was determined by measurements for central Pb+Pb 
reactions \cite{NA49EDHYP}, while for the C+C and Si+Si collisions $T$
was set to 200~MeV.  The rapidity spectra of \lab\ and \xim\ for Pb+Pb
reactions were modeled by single Gaussian distributions.  The flatter 
\lam\ rapidity spectra were instead approximated by the sum of two 
(Pb+Pb at 40\agev) or three (Pb+Pb at 158\agev) Gaussians, respectively.
For the C+C and Si+Si collisions the distributions for \lam\ and \lab\
were assumed to be constant as a function of rapidity.  The Geant~3.21 
package \cite{GEANT3} was used to track the generated particles and their 
decay products through the NA49 detector.  The TPC response, which takes 
into account all known detector effects, was simulated by software 
developed for the NA49 experiment.  The simulated signals were added to 
those of real events on the raw data level and subjected to the same 
reconstruction procedure as the experimental data.  By determining the 
fraction of the generated \lam, \lab\ and \xim\ which traverse the 
detector, survive the reconstruction, and pass all analysis cuts, the 
combined acceptances and efficiencies were derived.  The corresponding 
correction factors were calculated in bins of \pt, (\mtmzero), $y$, as 
well as centrality in the case of Pb+Pb collisions.  

Figure \ref{fig:eff_fd_corr}a summarizes the centrality 
dependence of the efficiencies, including acceptance.  While for \lam\ 
and \lab\ at 40\agev\ only a 30~\% variation can be observed, the 
difference between very peripheral and very central bins is a factor of 
3 at 158\agev.  
For the \lam\ also an alternative analysis strategy was employed: on top 
of the standard cuts, only those \lam\ were accepted whose daughter 
tracks lie entirely outside the high track density region (CUT-B in 
\cite{NA49EDHYP}).  This reduces the overall efficiency, but has the 
benefit that the track multiplicity effects are slightly reduced compared 
to the standard analysis (see dashed line in \Fi{fig:eff_fd_corr}a).  The 
influence of the detector occupancy on the efficiency is much smaller 
for C+C and Si+Si reactions at 158\agev.  It was found that the 
reduction of the efficiency due to other tracks is $\sim$~5~\% for 
Si+Si and negligible for C+C.  Therefore, it was only corrected for 
in the case of Si+Si reactions.  Due to the relaxed analysis cuts, 
the efficiencies are generally higher for the small systems than for 
minimum bias Pb+Pb reactions (\lam/\lab: $\sim$~55~\%, \xim: $\sim$~6~\%, 
both mid-rapidity) \cite{INGRID,MICHI}.

In addition to the directly produced particles, the measured yield of \lam\ 
and \lab\ contains also contributions from the decay of heavier 
hyperons.  The \lam\ (\lab) resulting from electromagnetic decays of 
\sig\ (\sib) cannot be separated from the directly produced ones via a
secondary vertex measurement.  Thus the presented yields always represent 
the sum $\lam + \sig$ ($\lab + \sib$).  The contribution to \lam\ 
(\lab) from weak decays, however, depends on the chosen analysis cuts, because 
these decay products originate from decay vertices with a sizable distance 
from the main interaction point.  Since the NA49 acceptance for \lam\ (\lab) 
favors those that decay at larger distances, the contribution of feed-down 
\lam\ (\lab) can be quite substantial.  Therefore, we calculated a 
correction for the feed-down from $\xim + \xizero$ ($\xip + \xib$) decays 
to the measured \lam\ (\lab) sample using the same simulation procedure as 
described above for the efficiency correction.  In this case a sample of 
\xim\ and \xizero\ (\xip\ and \xib) was generated as input.  The feed-down 
correction was then calculated in bins of \pt, (\mtmzero), $y$, as well as
reaction centrality, as the fraction of reconstructed \lam\ (\lab) which 
originate from $\xim + \xizero$ ($\xip + \xib$) decays and pass the same 
analysis cuts.  The \xim\ yields used as input to this procedure are based 
on the measurements presented in this publication, which were 
interpolated to all centrality bins.  The extrapolation of the \xip\ yields
to the different centrality bins was based on the yield measured in central 
Pb+Pb reactions \cite{NA49EDHYP}.  It was assumed that the centrality 
dependence of the \xip\ yields is the same as measured for the \xim.  In 
both cases the shapes of the input rapidity and \pt~distributions are a 
parametrization of the spectra measured for central collisions.  For the 
\xizero\ (\xib), which are not measured, the same shape of the phase space 
distributions as for the \xim\ was assumed.  Their yields are calculated 
by scaling those of \xim\ (\xip) by the \xizero/\xim\ (\xib/\xip) ratios 
taken from statistical model fits \cite{BECATTINI2}.  As an example, the 
\pt-integrated feed-down contribution around mid-rapidity is shown in 
\Fi{fig:eff_fd_corr}b.  Since the $\Xi$ yields have a
stronger centrality dependence than the $\Lambda$ yields, the feed-down
contribution changes with centrality.  In the case of C+C and Si+Si reactions 
all yields entering the calculation of the feed-down are based on statistical 
model fits \cite{BECATTINI2}.  The parameters for their phase space 
distributions are adjusted such as to interpolate between p+p and Pb+Pb 
reactions.  The corrections amount to 9~\% (10~\%) for \lam\ and 15~\% 
(20~\%) for \lab\ in C+C (Si+Si) \cite{INGRID}.  

\subsection{\label{sec:sys_err} Systematic errors}

%
\begin{table}[tbh]
\caption{\label{tab:syserr}
Summary of the systematic errors on the \dndy~values for minimum bias
Pb+Pb reactions.}
\begin{ruledtabular}
\begin{tabular}{llccccc}
     & \multicolumn{1}{l}{\ebeam}
     & \multicolumn{1}{l}{Bgnd.}
     & \multicolumn{1}{l}{Eff.}
     & \multicolumn{1}{l}{\pt~Extra-}
     & \multicolumn{1}{l}{Feed.}
     & \multicolumn{1}{l}{Quad.}      \\
     & \multicolumn{1}{l}{\agev}
     & \multicolumn{1}{l}{subtr.}
     & \multicolumn{1}{l}{corr.}
     & \multicolumn{1}{l}{polation}
     & \multicolumn{1}{l}{corr.}
     & \multicolumn{1}{l}{sum}        \\
                                                                                 \hline
       \lam &  40    &             &              &  --- &             & 11\%   \\
            & 158    & \rsbox{3\%} & \rsbox{10\%} &  3\% & \rsbox{3\%} & 11.5\% \\
                                                                                \hline
       \lab &  40    &             &              &  --- &             & 13\%   \\
            & 158    & \rsbox{3\%} & \rsbox{10\%} &  3\% & \rsbox{8\%} & 13.5\% \\
                                                                                \hline
       \xim & 40/158 &        3\%  &        10\%  &  3\% &  ---        & 11\%   \\
\end{tabular}
\end{ruledtabular}
\end{table}
%

The contributions to the systematic error of the \dndy~values measured in 
centrality selected Pb+Pb reactions are listed in \Ta{tab:syserr}.  The 
first two, the uncertainty of the background subtraction and the efficiency
correction, are identical to the ones determined for the analysis of the 
central Pb+Pb datasets \cite{NA49EDHYP}.  However, there is a difference 
in the contribution from the feed-down correction to the systematic 
error in the central Pb+Pb analysis.  The yields of \xim\, and particularly 
of \xip, are less precisely measured for non-central Pb+Pb reactions than 
for central ones.  While the feed-down contribution from \xim\ and \xizero\ 
to \lam\ can still be constrained reasonably well with the measurement 
presented here, the feed-down estimate from \xip\ and \xib\ to \lab\ relies 
on an extrapolation of the measurement in central collisions assuming the 
same centrality dependence.  By varying the $\Xi$ input yields inside the 
errors obtained for the central data set and assuming different centrality 
dependences (e.g. scaling with \nwound) the contribution to the systematic 
error was evaluated.  As a result a systematic error of 3~\% was assigned 
to the \lam, while for the \lab\ it is 8~\%.  Since the minimum bias data 
at 158\agev, and for \xim\ also at 40\agev, do not allow to measure the 
\pt-range down to $\pt = 0$~\gevc, an extrapolation has to be used.  By 
comparing the result for the standard fit (exponential, as shown in 
\Fis{fig:lm_lb_xi_mt_40}{fig:lm_lb_xi_mt_158}) to an extrapolation using 
a fit with a hydrodynamically inspired blast-wave model \cite{BLASTWAVE}, 
a systematic error of 3~\% was determined for these cases.  
Figure~\ref{fig:syserr} demonstrates the consistency of the systematic 
error estimated for the \lam\ yield with the spread of results obtained
from the two analysis strategies discussed above.

The extrapolations in rapidity, which are needed to extract the total 
multiplicities, introduce additional systematic errors.  The data allow 
to constrain the widths of the fits, as shown in 
\Fis{fig:lm_lb_y_40}{fig:lm_lb_y_158}, only to a certain extent which 
translates into an uncertainty of the extrapolation.  In the case of
\lam\ at 158\agev, the shapes of the rapidity distributions are not measured.  
Therefore a set of assumptions based on other measurements as discussed in 
\cite{NA49EDHYP} was used.  The shaded areas in \Fi{fig:lm_lb_y_158} 
represent the uncertainty of the extrapolations that were included in 
the systematic error.


\section{Results}

\subsection{Transverse mass spectra}

%
\begin{figure}[t]
\begin{center}
\includegraphics[width=\linewidth]{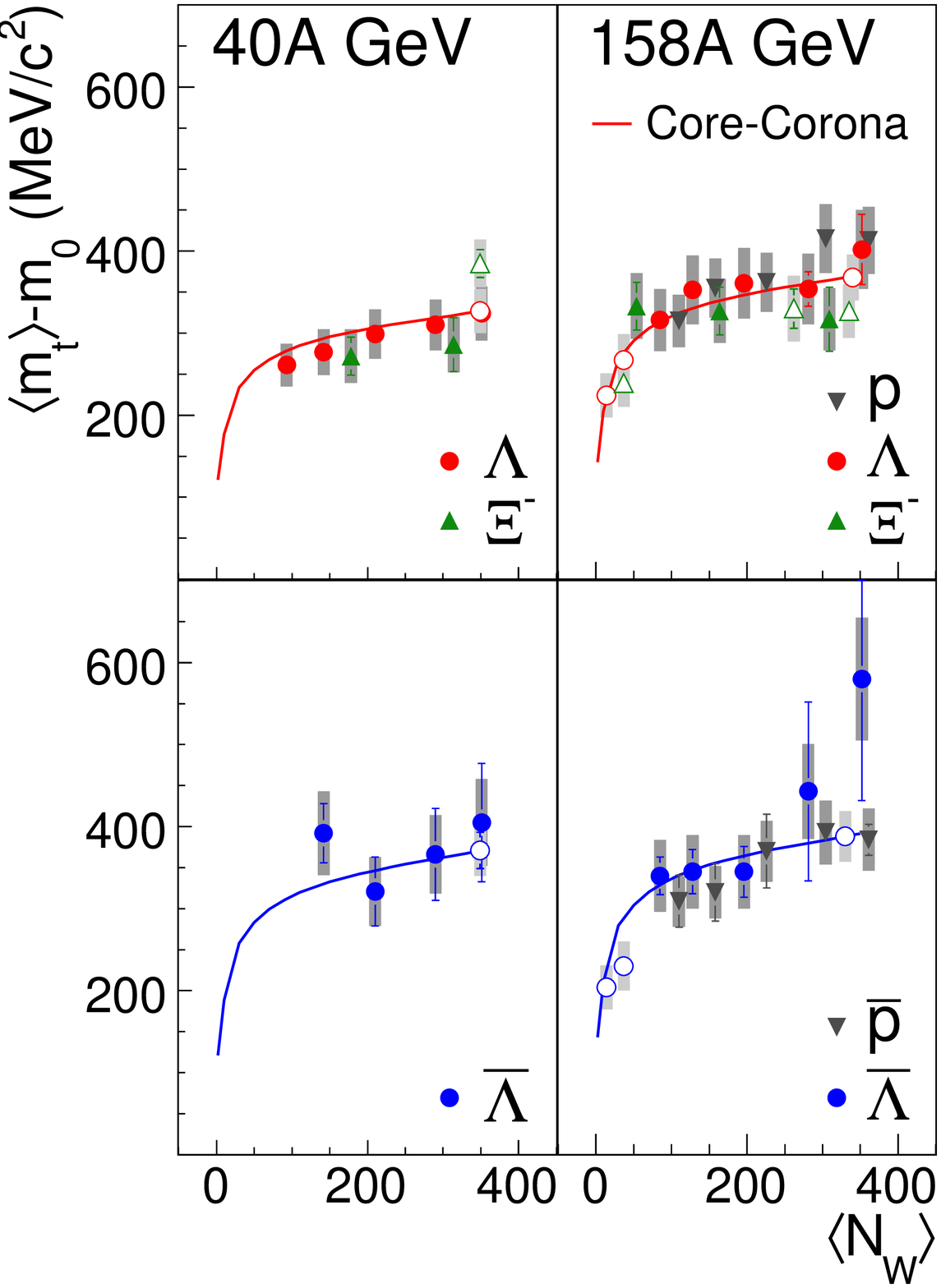}
\end{center}
\caption{(color online)
The \mtavg\ values at mid-rapidity (p/$\bar{\rm{p}}$: -0.5~$< y <$~-0.1, 
\lam/\lab: $|y| < 0.4$, \xim: $|y| < 0.5$) for Pb+Pb collisions at 40$A$ 
and 158\agev, as well as for near-central C+C and Si+Si reactions at 
158\agev.  The systematic errors are represented by the gray boxes.  The 
filled symbols correspond to the results obtained from the minimum bias 
data sets, while the open ones represent the (near-)central reaction 
systems.  The (anti-)proton data are taken from \cite{NA49SDPR1}.  Also 
shown are the results from a fit for \lam\ and \lab\ with the core-corona 
approach (solid lines).
}
\label{fig:meanmt_vs_nw} 
\end{figure} 
%

%
\begin{figure*}[htb]
\includegraphics[width=0.95\linewidth]{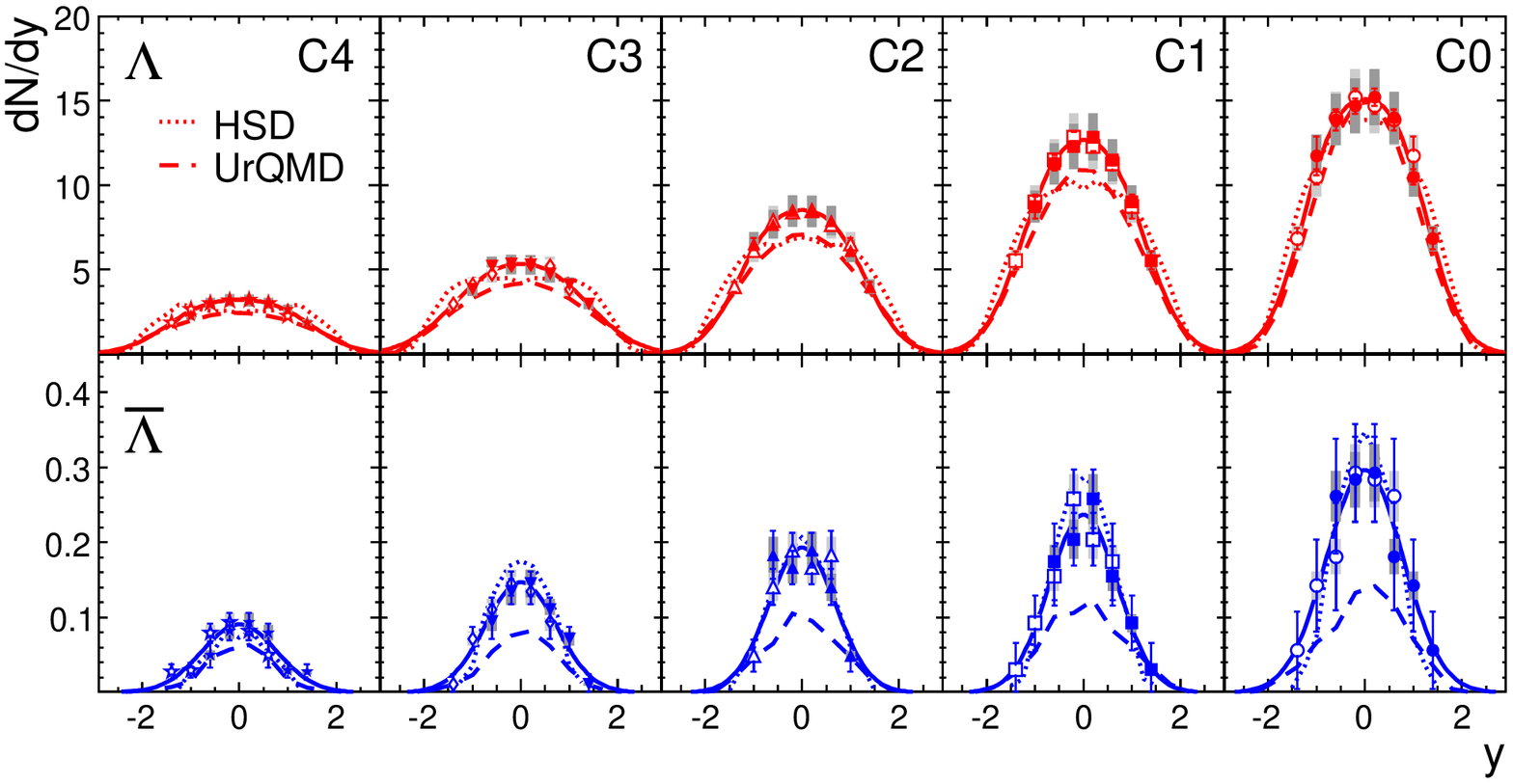}
\caption{(color online)
The rapidity spectra of \lam\ and \lab\ for Pb+Pb collisions at 40\agev\ 
for the 5 different centrality bins C0~--~C4.  The open symbols show 
data points reflected around mid-rapidity.  The systematic errors are
represented by the gray boxes.  Solid lines are fits to the data points,
used to extrapolate the measurements in order to extract total yields.  
Also included are calculations with the HSD model \cite{HSD1,HSD2,HSD3} 
(dotted lines) and the UrQMD2.3 model \cite{URQMD,URQMD23A,URQMD23B} 
(dashed lines).  
}
\label{fig:lm_lb_y_40} 
\end{figure*}
%

The transverse mass spectra of \lam, \lab, and \xim, measured around 
mid-rapidity (\lam/\lab: $|y| < 0.4$, \xim: $|y| < 0.5$), are shown for
different centrality classes of minimum bias Pb+Pb collisions at 40\agev\
in \Fi{fig:lm_lb_xi_mt_40} and at 158\agev\ in \Fi{fig:lm_lb_xi_mt_158}.
Also included in \Fi{fig:lm_lb_xi_mt_158} are the \mt~spectra for near-central 
C+C and Si+Si reactions at 158\agev.  The \mt~spectra were fitted by an 
exponential as defined in \Eq{eq:expo} in the transverse mass range 
$\mtmzero >  0.2$~\gevcc\ (Pb+Pb data) and 
$\mtmzero >  0.0$~\gevcc\ (C+C and Si+Si data).  The resulting inverse 
slope parameters $T$ are listed in 
\Tass{tab:summaryMB40}{tab:summaryMB158}{tab:summaryCN158}.  

For a model independent study of the energy dependence of \mt~spectra, the 
average transverse mass \mtavg\ was calculated.  To account for the 
unmeasured \mt~range two different parametrizations were employed to 
extrapolate: an exponential function (shown in 
\Fis{fig:lm_lb_xi_mt_40}{fig:lm_lb_xi_mt_158}) and a blast-wave function 
\cite{BLASTWAVE} (not shown).  Both provide a good description of the measured 
data.  An estimate of the systematic error is derived from the differences 
between the two approaches.  The resulting values for \mtavg\ are listed in 
\Tass{tab:summaryMB40}{tab:summaryMB158}{tab:summaryCN158}.

Figure~\ref{fig:meanmt_vs_nw} shows the dependence of \mtavg\ on \nwound\ for 
the hyperon data compared to previously published proton and antiproton 
results \cite{NA49SDPR1}.  The mass differences between the shown particle 
species are not very large and their \mtavg\ values agree within errors for 
each particular collision system.  However, there is a significant system-size 
dependence.  A strong increase is observed for \nwound\ below 60, whereas
above this region the values of \mtavg\ rise slowly with centrality in Pb+Pb 
collisions.

%
\begin{figure*}[htb]
\includegraphics[width=0.95\linewidth]{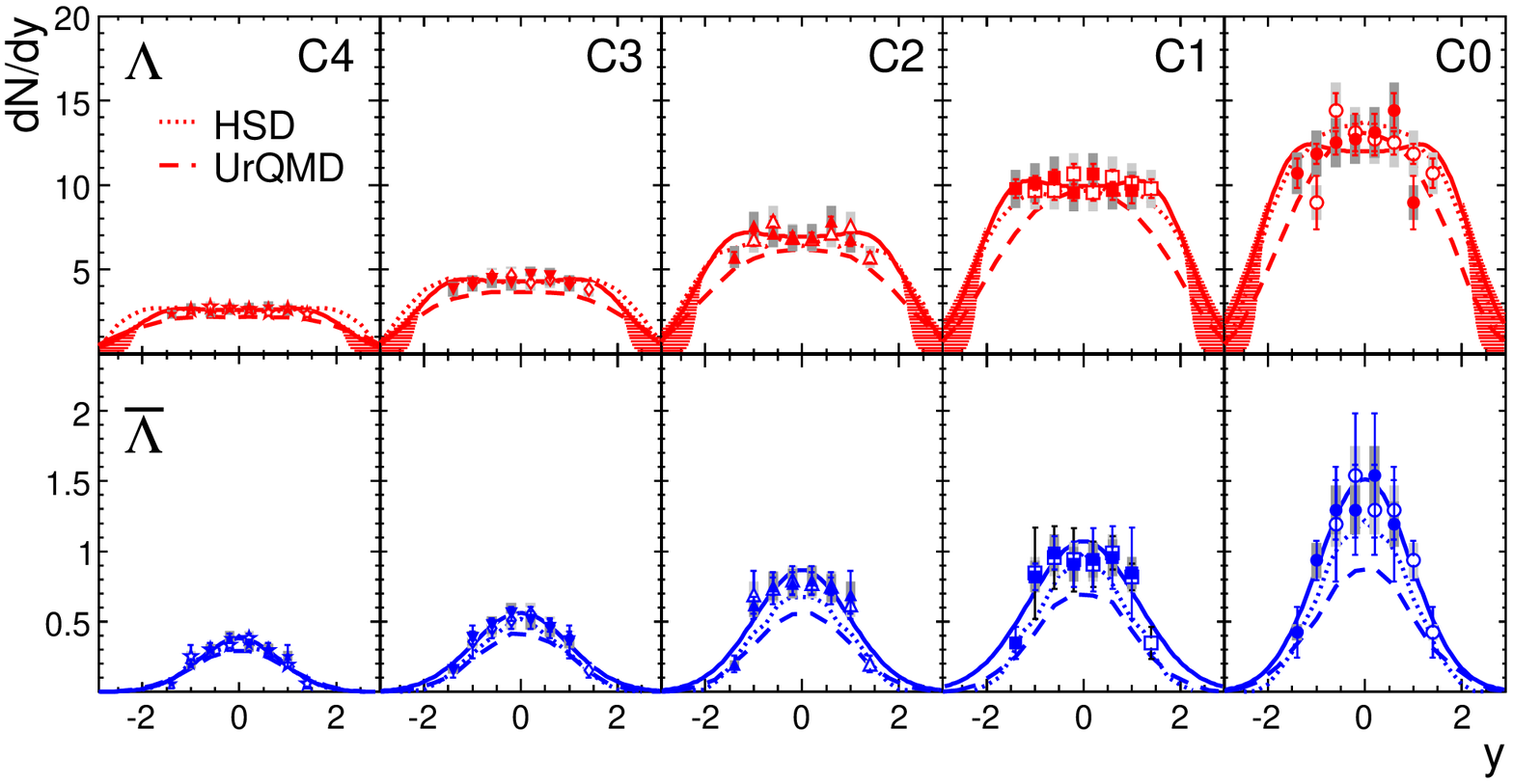}
\caption{(color online)
The rapidity spectra of \lam\ and \lab\ for Pb+Pb collisions at 158\agev\ 
in 5 different centrality bins C0~--~C4.  The open symbols show data 
points reflected around mid-rapidity.  The systematic errors are
represented by the gray boxes.  Solid lines are fits to the data points,
used to extrapolate the measurements in order to extract total yields.  
The shaded areas in the \lam\ spectra sketches the uncertainty due to the 
different extrapolations.  Also included are calculations with the HSD model 
\cite{HSD1,HSD2,HSD3} (dotted lines) and the UrQMD2.3 model 
\cite{URQMD,URQMD23A,URQMD23B} (dashed lines). 
}
\label{fig:lm_lb_y_158} 
\end{figure*}
%

\subsection{\label{sec:rap_spectra} Rapidity spectra}

%
\begin{figure}[t]
\begin{center}
\includegraphics[height=88mm]{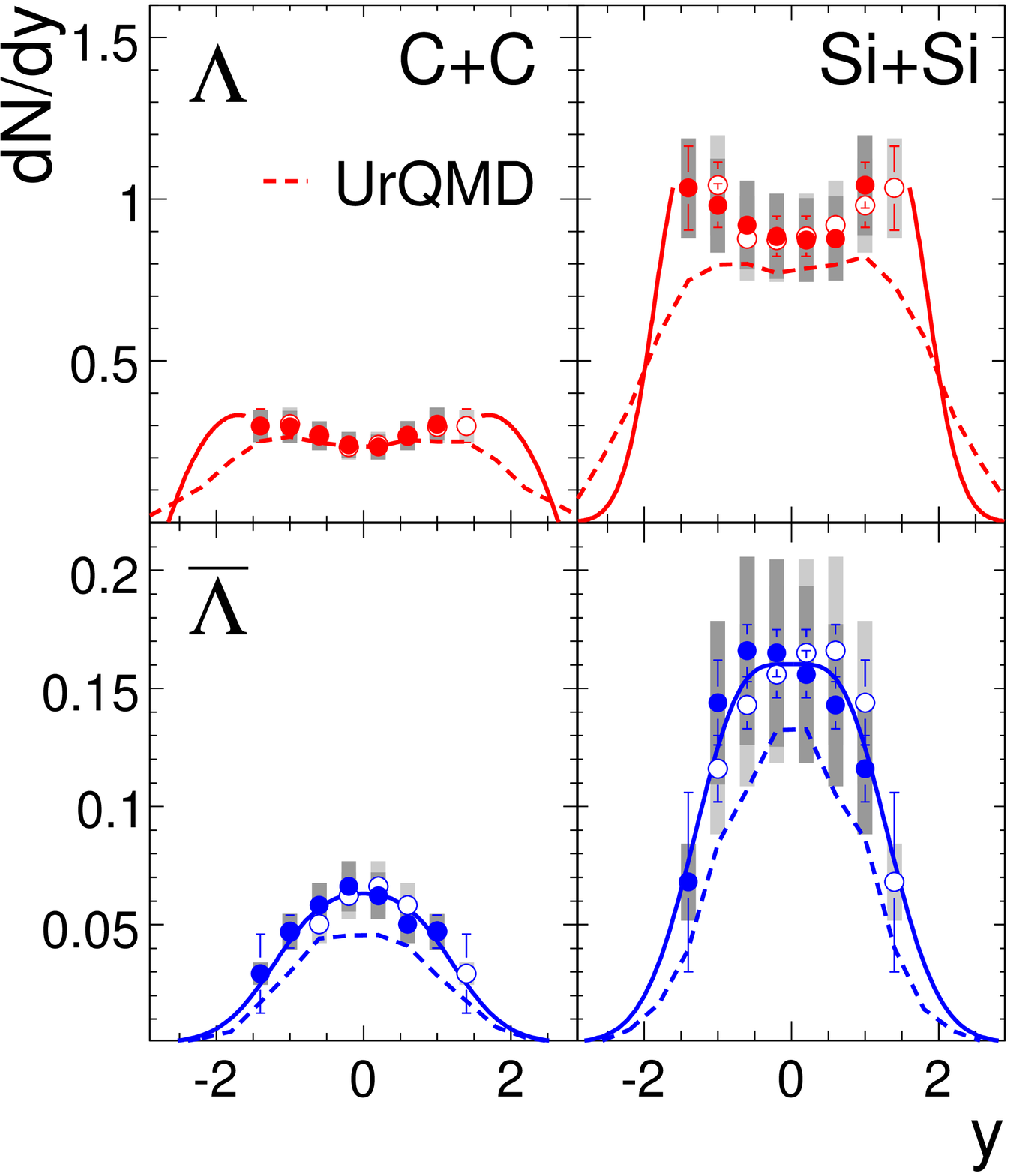}
\end{center}
\caption{(color online)
The rapidity spectra of \lam\ and \lab\ for near-central C+C and Si+Si 
collisions at 158\agev\ (the \lam\ spectra have already been published
in \cite{NA49CCSISI}).  Open symbols show data points reflected 
around mid-rapidity, while the systematic errors are represented by the 
gray boxes.  Solid lines are fits to the data points, used to extrapolate 
the measurements in order to extract total yields.  Also included are 
calculations with the UrQMD2.3 model \cite{URQMD,URQMD23A,URQMD23B} 
(dashed lines).
}
\label{fig:lm_lb_y_cc_sisi_158} 
\end{figure} 
%

%
\begin{figure}[t]
\begin{center}
\includegraphics[width=\linewidth]{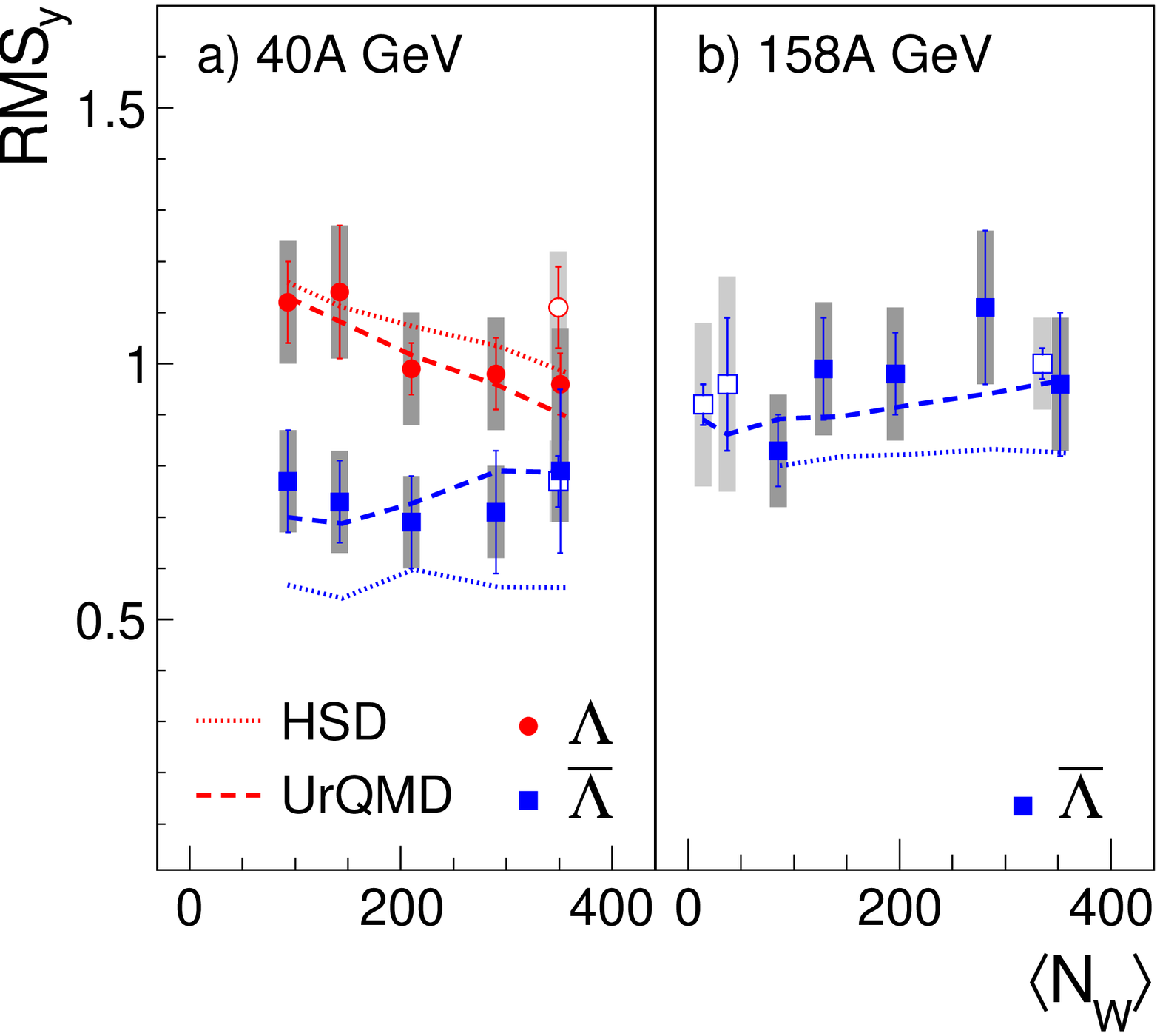}
\end{center}
\caption{(color online)
The RMS widths of the rapidity distributions $RMS_{\rb{y}}$ of \lam\ and \lab\ 
at 40$A$ and 158\agev\ as a function of the number of wounded nucleons 
\nwound.  The systematic errors are represented by the gray boxes.  Filled
symbols correspond to centrality selected data, obtained with a minimum bias
trigger, while the open ones represent the reaction systems measured with a
(near-)central trigger.  Also included are calculations with the HSD model 
\cite{HSD1,HSD2,HSD3} (dotted lines) and the UrQMD2.3 model 
\cite{URQMD,URQMD23A,URQMD23B} (dashed lines). 
}
\label{fig:rms_vs_nw}
\end{figure} 
%

The rapidity spectra of \lam\ and \lab\ for Pb+Pb collisions at 40$A$ and 
158\agev\ are summarized in \Fis{fig:lm_lb_y_40} {fig:lm_lb_y_158}.
For \xim\ the statistics of the minimum bias datasets was not sufficient
to extract rapidity spectra.  
While the \lam\ distributions at 40\agev\ and the \lab\ distributions at 
40$A$ and 158\agev\ have a Gaussian shape at all centralities, the \lam\ 
distributions at 158\agev\ are rather flat over the measured rapidity range, 
similar to what has been observed for central Pb+Pb reactions \cite{NA49EDHYP}.  
The \lam\ and \lab\ spectra for the near-central C+C and Si+Si collisions are 
shown in \Fi{fig:lm_lb_y_cc_sisi_158}.  Also here the rapidity distributions 
for \lam\ are relatively flat with an indication for a minimum at mid-rapidity, 
which appears to be even more pronounced than in the case of the Pb+Pb data.  
The corresponding rapidity densities \dndy\ around mid-rapidity for all data 
sets are listed in \Tass{tab:summaryMB40}{tab:summaryMB158}{tab:summaryCN158}.

The determination of total multiplicities requires an extrapolation into the
unmeasured $y$ regions.  For this purpose the \lam\ distributions at 40\agev\ 
were fitted with the sum of two Gauss functions of equal width $\sigma$ which
are displaced symmetrically by $s$ with respect to mid-rapidity:
\begin{equation}
  \label{eq:twogaus}
  \!\! 
  \frac{\der N}{\der y} 
  \propto 
  \exp \left\{ - \frac{(y - s)^{2}}{2 \sigma^{2}} \right\} +
  \exp \left\{ - \frac{(y + s)^{2}}{2 \sigma^{2}} \right\} .
\end{equation}
At 158\agev, the data do not allow to determine the shape of the \lam\
rapidity spectra outside the plateau region around mid-rapidity.  We therefore
use the same assumptions on the spectral shape that have been applied
to the central Pb+Pb data at 158\agev\ \cite{NA49EDHYP} for all centrality
bins in order to extract total multiplicities.  This, of course, assumes
that the widths of the rapidity distributions do not change substantially
with centrality.  For the \lab\ spectra a single Gaussian provides a 
reasonable fit at both beam energies.

The fitted $RMS_{\rb{y}}$ values are tabulated in 
\Tass{tab:summaryMB40}{tab:summaryMB158}{tab:summaryCN158}.
Figure~\ref{fig:rms_vs_nw} displays the system-size dependence of 
$RMS_{\rb{y}}$.  While for \lam\ at 40\agev\ an indication for a decrease 
of the widths with increasing centrality can be seen, no significant 
system-size dependence is observed for \lab\ at both beam energies.

\subsection{Particle yields}

%
\begin{figure}[t]
\begin{center}
\includegraphics[width=\linewidth]{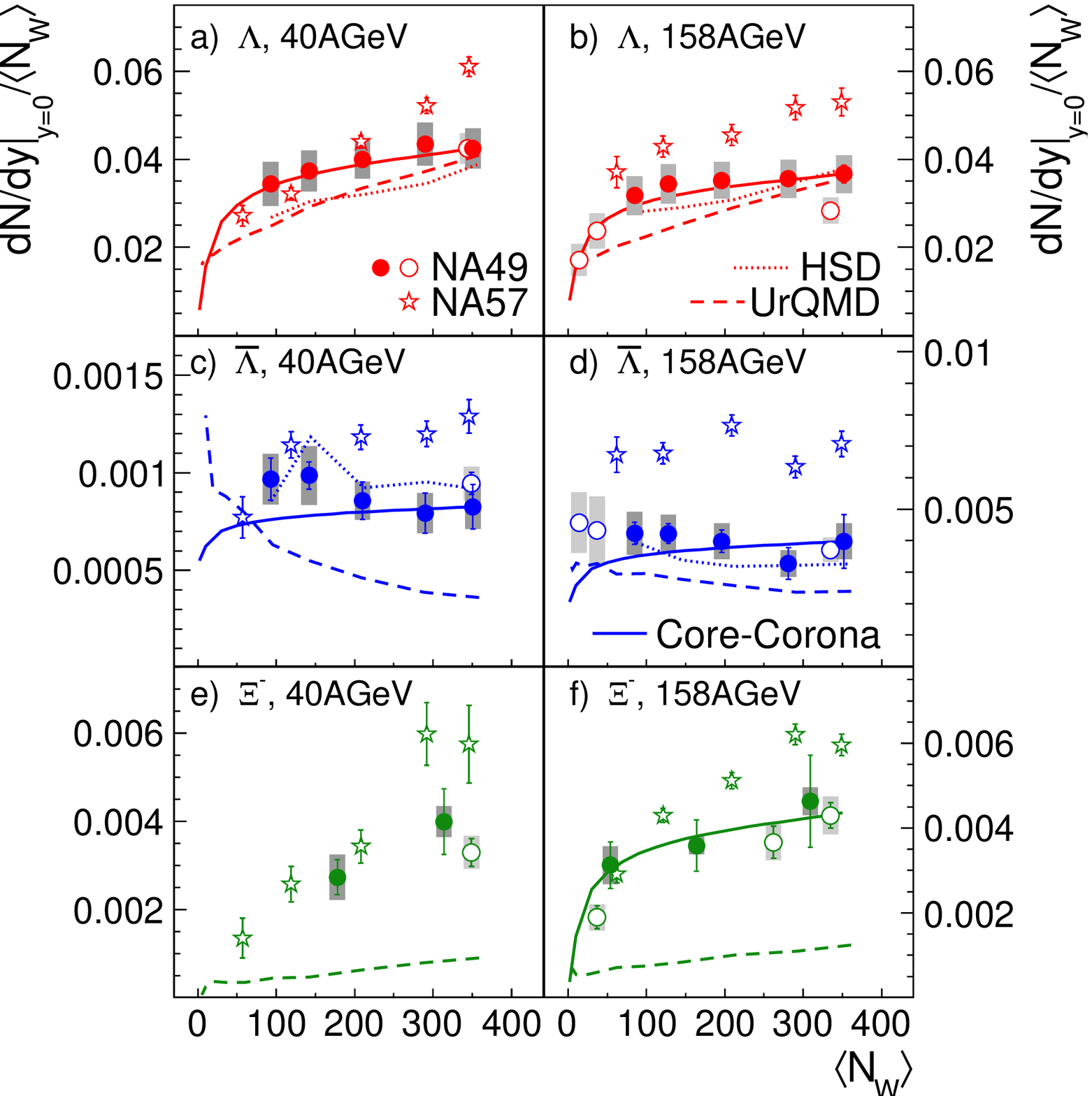}
\end{center}
\caption{(color online)
The rapidity densities \dndy\ divided by the average number of wounded 
nucleons \nwound\ of \lam, \lab, and \xim\ at mid-rapidity (\lam/\lab: 
$|y| < 0.4$, \xim: $|y| < 0.5$) for Pb+Pb collisions at 40$A$ and 158\agev, 
as well as for near-central C+C and Si+Si reactions at 158\agev, as a 
function of \nwound.  The systematic errors are represented by the gray 
boxes.  Filled symbols correspond to the minimum bias trigger, while the 
open ones represent the online triggered (near-)central reaction systems.  
Also shown are data of the NA57 collaboration \cite{NA57HY40,NA57HY158} 
(open stars) and calculations with the HSD model \cite{HSD1,HSD2,HSD3} 
(dotted lines), the UrQMD2.3 model \cite{URQMD,URQMD23A,URQMD23B} (dashed 
lines), and the core-corona approach (solid lines).
}
\label{fig:dndy_vs_nw}
\end{figure}
%

%
\begin{figure}[t]
\begin{center}
\includegraphics[width=\linewidth]{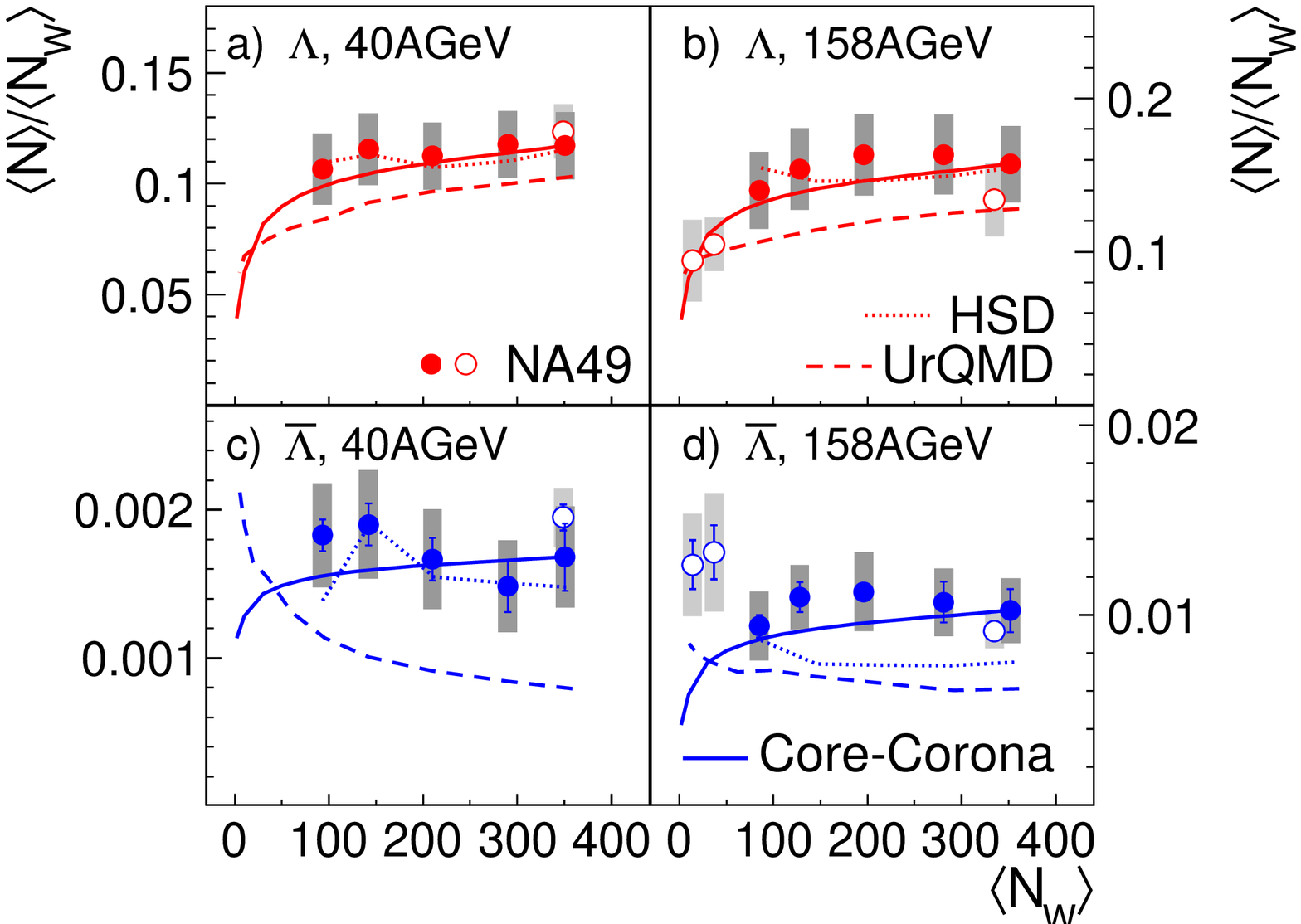}
\end{center}
\caption{(color online)
The total multiplicities \navg\ divided by the average number of 
wounded nucleons \nwound\ of \lam\ and \lab\ for Pb+Pb collisions at 
40$A$ and 158\agev, as well as for near-central C+C and Si+Si reactions 
at 158\agev, as a function of \nwound.  The systematic errors are 
represented by the gray boxes.  Filled symbols correspond to the 
minimum bias trigger, while the open ones represent online triggered 
(near-)central reaction systems.  Also shown are calculations with the 
HSD model \cite{HSD1,HSD2,HSD3} (dotted lines), the UrQMD2.3 model 
\cite{URQMD,URQMD23A,URQMD23B} (dashed lines), and the core-corona 
approach (solid lines).
}
\label{fig:yield_vs_nw}
\end{figure}
%

Figure~\ref{fig:dndy_vs_nw} shows the system-size dependences of the
rapidity densities \dndy\ at mid-rapidity for \lam, \lab, and \xim, 
divided by the average number of wounded nucleons \nwound.  For 
$\nwound > 60$, \dndy/\nwound\ of \lam\ is almost independent of the 
system size, while for smaller systems at 158\agev, corresponding to
a \nwound\ range not covered by Pb+Pb collisions, a significant rise 
with \nwound\ is observed.  However, one should keep in mind that a 
direct comparison of near-central C+C and Si+Si reactions to peripheral 
Pb+Pb collisions is complicated by the fact that the surface to volume 
ratio of these reaction systems is different.  In the case of \lab, 
this ratio seems to be independent from the system size even for very 
small systems.  The \xim, on the other hand, exhibits a weak \nwound\ 
dependence.  For comparison, data of the NA57 collaboration are also 
included in \Fi{fig:dndy_vs_nw}.  Generally, the \dndy~values of NA57 
are higher for non-central Pb+Pb collisions, similar to what has already 
been found for central Pb+Pb \cite{NA49EDHYP}.  It appears, however, 
that the discrepancy becomes smaller for peripheral collisions.

The total multiplicities \navg\ of \lam\ and \lab\ at 40$A$ and 158\agev, 
as determined from extrapolations of the rapidity spectra shown in 
\Fiss{fig:lm_lb_y_40}{fig:lm_lb_y_158}{fig:lm_lb_y_cc_sisi_158} normalized
by \nwound\ are summarized in \Fi{fig:yield_vs_nw}.  A similar picture 
emerges as for the mid-rapidity \dndy\ values.  In the range $\nwound > 60$ 
the ratio \navg/\nwound\ is independent of the system size.  Towards 
smaller system sizes, covered only by C+C and Si+Si collisions at 158\agev,
a significant decrease for \lam\ is observed, while in case of the \lab\
this ratio remains constant throughout.


\section{Discussion}

\subsection{Comparison to transport models}

Transport models allow to study several effects that may influence the 
system size dependence of strange particle production, e.g. multi-meson 
fusion processes, absorption of anti-baryons in the fireball, and the 
evolution of the longitudinal distribution of baryon number.

Multi-meson fusion processes are a possible mechanism to enhance the 
production of anti-baryons \cite{HSD4} and may therefore be important for 
reaching statistical equilibrium yields of multi-strange anti-baryons 
\cite{CGREINER}.  The HSD model \cite{HSD1,HSD2,HSD3} offers the 
possibility to include these fusion processes.  The 
Figures~\ref{fig:dndy_vs_nw} and \ref{fig:yield_vs_nw} show comparisons 
of HSD and UrQMD2.3~\cite{URQMD,URQMD23A,URQMD23B} to the measured yields 
at mid-rapidity and to the total yields.  For \lam\ the predictions of 
both models are close to the data.  However, the system-size dependence, 
especially for the total yields, seems to be better described by HSD.  
The multi-meson fusion processes are naturally most important for \lab.  
This explains why the spectra and yields for \lab\ in Pb+Pb predicted
by HSD are higher than those from UrQMD2.3, which does not feature these 
processes. HSD gives thus a better description of the measurements. The 
yield of \xim\ is underestimated by UrQMD2.3 by factors of 2~--~3 for 
all systems.  No HSD calculations for the \xim\ are available yet.  

The system-size dependence of anti-baryon yields should also be affected
by their possible absorption in the surrounding dense matter of the fireball.  
In this case one would expect the measured \lab\ yield per wounded nucleon 
to go down when comparing the small C+C and Si+Si systems with central Pb+Pb 
collisions.  In fact, the data on \navg/\nwound\ for \lab\ at 158\agev\ seem 
to exhibit the expected tendency to decrease from C+C towards Pb+Pb collisions 
(see \Fi{fig:yield_vs_nw}d), quite in contrast to the \lam, where \navg/\nwound\ 
is rather increasing in the region $\nwound < 60$ (see \Fi{fig:yield_vs_nw}b).  
A similar behavior is predicted by UrQMD2.3, where it is, however, stronger 
at 40\agev\ than at 158\agev.   But due to the size of the systematic error 
of the measurements, no final conclusion can be made whether \lab\ production 
is really affected by absorbtion.

Figures~\ref{fig:lm_lb_y_40}, \ref{fig:lm_lb_y_158}, and 
\ref{fig:lm_lb_y_cc_sisi_158} include a comparison of the transport model
predictions to the measured rapidity distributions of \lam\ and \lab.  In 
the case of \lab\ the predicted widths of the rapidity distributions from 
both models, UrQMD2.3 and HSD, fit the data for all studied systems and 
energies reasonably well (see \Fi{fig:rms_vs_nw}).  \lam\ rapidity spectra,
which are sensitive to the final distribution of baryon number, exhibit a 
significant dependence of their shape on system size.  Also here the 
agreement to the models is fairly good at both energies, even though 
UrQMD2.3 predicts a Gaussian shaped distribution at 158\agev, while the 
data would rather suggest a plateau inside the measured region.  HSD, on 
the other hand, describes this flat shape relatively well.  Similar 
observations have been made in the case of proton rapidity distributions 
in minimum bias Pb+Pb reactions at 158\agev, where HSD also gives a better 
agreement with the observed flat proton spectra than UrQMD2.3 
\cite{NA49SDPR2}.  The reason for this difference lies in a different 
assumption on when a nucleon is allowed to interact again after its first 
collision.  On top of a formation time of $\tau = 0.8$~fm/$c$, which is 
implemented in both models, HSD requires that the local energy density 
falls below 1~GeV/fm$^3$, which is considered as the critical energy 
density for a phase transition to a QGP.  Thus, the data would suggest 
that this additional criterion is needed to properly describe the 
redistribution of baryon number in longitudinal phase space due to
stopping.

\subsection{Core-Corona approach}

In order to compare the core-corona approach with the data presented here,
we generalize the prescription given in \cite{WERNER2} and parametrize 
the system-size dependence of any observable $X$ by:
\begin{eqnarray}
\label{eq:core_corona}
X(\nwound) & = & \nwound \: [ f(\nwound)      \:X_{\rb{core}} \\
           &   & \:\:\:\:\:\:\:\:\:\:\: + \: (1 - f(\nwound)) \:X_{\rb{corona}} ] \nonumber
\end{eqnarray}
The quantity $X$ can either be the average transverse mass \mtavg, the 
rapidity density \dndy, or the total multiplicity \navg.  The function 
$f(\nwound)$ is here defined as the fraction of all participating nucleons, 
which interact more than once, and can therefore be attributed to the core 
region.  Since the corona should behave like independent nucleon--nucleon 
collisions, the quantity $X_{\rb{corona}}$ corresponds to results of 
measurements in p+p collisions.  Thus, the function $f(\nwound)$ provides 
a natural interpolation between p+p and Pb+Pb reactions.  We use values for 
$f(\nwound)$ (see \Ta{tab:mbdatasets}), that have been calculated within a 
Glauber approach for Pb+Pb collisions at 158\agev\ and have also been used 
in the toy model comparison discussed in \cite{WERNER2}.  Since the 
nucleon--nucleon cross section changes only slightly between 40 and 158~GeV 
beam energy, we use the same values of $f(\nwound)$ for the comparison to the 
40\agev\ data.  It should be noted, though, that the direct comparison of the 
curves shown here to semi-central C+C and Si+Si collisions is not entirely 
correct, since their surface to volume ratio is different from that in Pb+Pb 
collisions. This, in principle, would require a calculation of $f({\nwound})$ 
specifically for these reaction systems.  More insight could also be gained 
by studying the smaller systems in several centrality bins, similar to the 
study of Cu+Cu in \cite{TIMMINS}.  However, our available statistics for C+C 
and Si+Si unfortunately does not allow this.

Based on the above recipe, the system-size dependence of \mtavg\ for \lam\ and
\lab\ was constructed (solid lines in \Fi{fig:meanmt_vs_nw}).  The \mtavg\ 
values for \lam\ in p+p collisions are based on an interpolation of p+p data 
measured at various beam energies \cite{PPLAM}.  The p+p value for \lab\ was 
assumed to be the same as for \lam, since not enough data is available to do
the extrapolation.  The core contributions to \mtavg\ were adjusted to the 
measurements for central Pb+Pb collisions.  In fact, the model provides a 
reasonable description of the measured system-size dependence in all cases.  

Similarly, the system-size dependence of \dndy/\nwound\ and \navg/\nwound\ can 
be predicted using the core-corona approach as given by \Eq{eq:core_corona}.  
The solid lines in \Fis{fig:dndy_vs_nw}{fig:yield_vs_nw} are based on the same 
function $f(\nwound)$ as has been used for \mtavg.  Here, $X_{\rb{corona}}$ is 
adjusted to the yields derived from an interpolation of \lam\ and \lab\ yields 
measured in p+p collisions at different beam energies \cite{MAREKPP}.  For the 
\xim\ at 158\agev\ a preliminary p+p measurement by NA49 was used \cite{TANJAQM}, 
while for 40\agev\ no p+p input is available so that no comparison to \xim\ is 
possible at this energy.  $X_{\rb{core}}$ is defined in all cases by the measured 
\dndy, resp. \navg, for central Pb+Pb collisions.  The agreement is good for the 
yields of \lam\ and \xim\ (see: \Fis{fig:dndy_vs_nw}{fig:yield_vs_nw}).  However, 
for \lab\ at 158\agev\ the yields measured in C+C and Si+Si collisions are at 
the same level as for Pb+Pb reactions.  This behaviour cannot be fitted 
by the core-corona approach and would therefore indicate that also other 
mechanisms, such as \lab-absorption, need to be taken into account to arrive at 
a proper description of the system-size dependence.  


\section{Summary}

A measurement of \lam, \lab, and \xim\ production in centrality selected
Pb+Pb collisions at 40$A$ and 158\agev\ and in near-central C+C and Si+Si
collisions at 158\agev\ is presented.  The first moments of the transverse 
mass spectra (\mtavg) exhibit only a weak system-size dependence for 
$\nwound > 60$, while for the small systems a rapid rise of \mtavg\ with 
increasing system size is observed.  The rapidity distributions of \lam\ 
at 40\agev\ and of \lab\ at 40$A$ and 158\agev\ have a Gaussian shape.
For \lam\ at 158\agev\ the rapidity spectra are rather flat in the measured 
region $-1.6 < y < 1.2$.  Generally, no pronounced system-size dependence 
of the widths of the rapidity distributions is observed.  Only the \lam\ 
spectra at 40\agev\ might show some indication for a slight narrowing with 
increasing centrality.  The measured \dndy/\nwound\ and \navg/\nwound\ 
values rise rapidly with system size for small systems ($\nwound < 60$) 
and do not change very much any more from then on.  The core-corona 
approach describes the system-size dependence of \mtavg\ for all particle 
species discussed here.  It also reproduces the system-size dependence of 
the mid-rapidity \dndy~values and of the total multiplicities of \lam\ 
and \xim.  However, the \lab\ spectra measured for C+C and Si+Si collisions 
at 158\agev\ suggest a flatter system-size dependence than expected in the 
core-corona picture.  Generally, the results of the hadronic transport 
models UrQMD2.3 and HSD for \lam\ and \lab\ are close to the data, with 
the exception of the underprediction of the absolute \lab\ yields at both
energies by UrQMD2.3.  But both models predict a system-size dependence of 
the \lab\ total multiplicity similar to the measurement.  This might 
indicate that absorption of \lab\ in the dense hadronic medium, which is 
taken into account in the hadronic transport models, has a visible effect.
However, these models are not able to describe the production of baryons with 
multiple strangeness.  UrQMD2.3, for example, underestimates the yields of 
\xim\ to a large extent (factor 2~--~3).  The HSD model seems to provide a 
better description of the \lam\ rapidity spectra than UrQMD2.3 due to an 
improved implementation of the stopping mechanism.


\begin{acknowledgments}
This work was supported by the US Department of Energy
Grant DE-FG03-97ER41020/A000,
the Bundesministerium f\"{u}r Bildung und Forschung, Germany (06F137), 
the Virtual Institute VI-146 of Helmholtz Gemeinschaft, Germany,
the Hungarian Scientific Research Foundation (T032648, T032293, T043514),
the Hungarian National Science Foundation, OTKA, (F034707),
the Polish-German Foundation, 
the Polish Ministry of Science and Higher Education (1 P03B 006 30,
1 P03B 127 30, 0297/B/H03/2007/33, N N202 078735), 
the Korea Research Foundation (KRF-2007-313-C00175) 
and the Bulgarian National Science Fund (Ph-09/05).
\end{acknowledgments}




\appendix
\section{Tables}

%
\begin{table*}[p]
\caption{
The rapidity densities at mid-rapidity (\lam/\lab: $|y| < 0.4$, \xim: $|y| < 0.5$),
the total multiplicities \navg, the RMS widths of the rapidity distributions 
$RMS_{\rb{y}}$ calculated from the fits shown in \Fi{fig:lm_lb_y_40}, the average
transverse masses \mtavg, and the inverse slope parameters $T$ for Pb+Pb 
collisions at 40\agev.  The first error is statistical, the second systematic.
}
\begin{ruledtabular}
\begin{tabular}{llllllll}
\label{tab:summaryMB40} 
     & Centrality
     & \nwound
     & \dndy
     & \navg
     & $RMS_{\rb{y}}$
     & \mtavg
     & $T$            \\
     & class
     &
     &
     &
     &
     & (\mevcc)
     & (MeV)          \\
\hline
\lam &    0 & 351$\pm$3 & 14.9$\pm$0.3$\pm$1.6    & 41.1$\pm$0.8$\pm$5.3    & 0.96$\pm$0.06$\pm$0.11 & 324$\pm$10$\pm$33 & 268$\pm$ 7$\pm$16 \\
     &    1 & 290$\pm$4 & 12.6$\pm$0.2$\pm$1.4    & 34.1$\pm$0.6$\pm$4.4    & 0.98$\pm$0.07$\pm$0.11 & 310$\pm$ 8$\pm$31 & 256$\pm$ 4$\pm$15 \\
     &    2 & 210$\pm$6 &  8.4$\pm$0.1$\pm$0.9    & 23.6$\pm$0.3$\pm$3.1    & 0.99$\pm$0.05$\pm$0.11 & 299$\pm$ 6$\pm$30 & 247$\pm$ 3$\pm$15 \\
     &    3 & 142$\pm$8 &  5.3$\pm$0.1$\pm$0.6    & 16.4$\pm$0.2$\pm$2.1    & 1.14$\pm$0.13$\pm$0.13 & 277$\pm$ 6$\pm$28 & 230$\pm$ 4$\pm$14 \\
     &    4 &  93$\pm$7 &  3.2$\pm$0.05$\pm$0.4   &  9.9$\pm$0.1$\pm$1.3    & 1.12$\pm$0.08$\pm$0.12 & 261$\pm$ 6$\pm$26 & 220$\pm$ 5$\pm$13 \\
\hline
\lab &    0 & 351$\pm$3 &  0.29$\pm$0.04$\pm$0.04 &  0.59$\pm$0.08$\pm$0.12 & 0.79$\pm$0.16$\pm$0.10 & 405$\pm$72$\pm$53 & 325$\pm$81$\pm$32 \\
     &    1 & 290$\pm$4 &  0.23$\pm$0.03$\pm$0.03 &  0.43$\pm$0.05$\pm$0.09 & 0.71$\pm$0.12$\pm$0.09 & 366$\pm$56$\pm$48 & 299$\pm$40$\pm$30 \\
     &    2 & 210$\pm$6 &  0.18$\pm$0.02$\pm$0.02 &  0.35$\pm$0.03$\pm$0.07 & 0.69$\pm$0.09$\pm$0.09 & 321$\pm$42$\pm$42 & 276$\pm$38$\pm$28 \\
     &    3 & 142$\pm$8 &  0.14$\pm$0.01$\pm$0.02 &  0.27$\pm$0.02$\pm$0.05 & 0.73$\pm$0.08$\pm$0.10 & 392$\pm$36$\pm$51 & 346$\pm$69$\pm$35 \\
     &    4 &  93$\pm$7 &  0.09$\pm$0.01$\pm$0.01 &  0.17$\pm$0.01$\pm$0.03 & 0.77$\pm$0.10$\pm$0.10 & ---               & ---               \\
\hline
\xim & 0--1 & 314$\pm$4 &  1.25$\pm$0.23$\pm$0.14 & ---                     & ---                    & 286$\pm$33$\pm$34 & 232$\pm$19$\pm$14 \\
     & 2--3 & 178$\pm$8 &  0.49$\pm$0.07$\pm$0.05 & ---                     & ---                    & 272$\pm$23$\pm$33 & 233$\pm$18$\pm$14 \\
\end{tabular}
\end{ruledtabular}
%
%
%
%
%
%
%
%

%
\caption{
The rapidity densities at mid-rapidity (\lam/\lab: $|y| < 0.4$, \xim: $|y| < 0.5$),
the total multiplicities \navg, the RMS widths of the rapidity distributions 
$RMS_{\rb{y}}$ calculated from the fits shown in \Fi{fig:lm_lb_y_158}, the average
transverse masses \mtavg, and the inverse slope parameters $T$ for Pb+Pb 
collisions at 158\agev.  The first error is statistical, the second systematic.
}
\begin{ruledtabular}
\begin{tabular}{llllllll}
\label{tab:summaryMB158}
     & Centrality
     & \nwound
     & \dndy
     & \navg
     & $RMS_{\rb{y}}$
     & \mtavg
     & $T$            \\
     & class
     & 
     & 
     & 
     & 
     & (\mevcc)
     & (MeV)          \\
\hline
\lam &    0 & 352$\pm$3 & 12.9$\pm$0.7$\pm$1.5    & 55.3$\pm$1.8$\pm$8.8   & ---                    & 402$\pm$43$\pm$48 & 346$\pm$34$\pm$21 \\
     &    1 & 281$\pm$4 & 10.0$\pm$0.4$\pm$1.2    & 45.9$\pm$1.0$\pm$7.3   & ---                    & 354$\pm$21$\pm$43 & 296$\pm$14$\pm$18 \\
     &    2 & 196$\pm$6 &  6.9$\pm$0.2$\pm$0.8    & 32.0$\pm$0.5$\pm$5.1   & ---                    & 361$\pm$16$\pm$43 & 307$\pm$11$\pm$18 \\
     &    3 & 128$\pm$8 &  4.4$\pm$0.1$\pm$0.5    & 19.7$\pm$0.3$\pm$3.2   & ---                    & 353$\pm$15$\pm$42 & 303$\pm$12$\pm$18 \\
     &    4 &  85$\pm$7 &  2.7$\pm$0.1$\pm$0.3    & 11.9$\pm$0.2$\pm$1.9   & ---                    & 316$\pm$14$\pm$38 & 274$\pm$15$\pm$16 \\
\hline
\lab &    0 & 352$\pm$3 &  1.4$\pm$0.3$\pm$0.2    & 3.6$\pm$0.4$\pm$0.6    & 0.96$\pm$0.14$\pm$0.13 & 580$\pm$148$\pm$75 & 507$\pm$211$\pm$51 \\
     &    1 & 281$\pm$4 &  0.92$\pm$0.14$\pm$0.12 & 3.0$\pm$0.3$\pm$0.5    & 1.11$\pm$0.15$\pm$0.15 & 443$\pm$109$\pm$58 & 372$\pm$62$\pm$37 \\
     &    2 & 196$\pm$6 &  0.78$\pm$0.07$\pm$0.11 & 2.2$\pm$0.1$\pm$0.4    & 0.98$\pm$0.08$\pm$0.13 & 345$\pm$31$\pm$45 & 296$\pm$27$\pm$30 \\
     &    3 & 128$\pm$8 &  0.54$\pm$0.04$\pm$0.07 & 1.4$\pm$0.1$\pm$0.2    & 0.99$\pm$0.10$\pm$0.13 & 345$\pm$27$\pm$45 & 302$\pm$34$\pm$30 \\
     &    4 &  85$\pm$7 &  0.36$\pm$0.03$\pm$0.05 & 0.8$\pm$0.05$\pm$0.14  & 0.83$\pm$0.07$\pm$0.11 & 340$\pm$23$\pm$44 & 309$\pm$50$\pm$31 \\
\hline
\xim & 0--1 & 309$\pm$4 &  1.43$\pm$0.33$\pm$0.16 & ---                    & ---                    & 317$\pm$39$\pm$38 & 244$\pm$41$\pm$15 \\
     & 2--3 & 164$\pm$8 &  0.59$\pm$0.10$\pm$0.06 & ---                    & ---                    & 327$\pm$29$\pm$39 & 264$\pm$39$\pm$16 \\
     & 4--5 &  54$\pm$7 &  0.17$\pm$0.03$\pm$0.02 & ---                    & ---                    & 333$\pm$29$\pm$40 & 261$\pm$37$\pm$16 \\
     & 0--2 & 262$\pm$4 &  0.96$\pm$0.10$\pm$0.11 & ---                    & ---                    & 330$\pm$24$\pm$40 & 263$\pm$14$\pm$16 \\
\end{tabular}
\end{ruledtabular}
%
%
%
%
%
%
%
%

%
\caption{
The rapidity densities at mid-rapidity (\lam/\lab: $|y| < 0.4$, \xim: $|y| < 0.5$),
the total multiplicities \navg, the RMS widths of the rapidity distributions 
$RMS_{\rb{y}}$ calculated from the fits shown in \Fi{fig:lm_lb_y_cc_sisi_158}, the 
average transverse masses \mtavg, and the inverse slope parameters $T$ for 
near-central C+C and Si+Si collisions at 158\agev.  The first error is statistical, 
the second systematic.
}
\begin{ruledtabular}
\begin{tabular}{llllllll}
\label{tab:summaryCN158}
     & Reaction
     & \nwound
     & \dndy
     & \navg
     & $RMS_{\rb{y}}$
     & \mtavg
     & $T$            \\
     & system
     &
     &
     &
     &
     & (\mevcc)
     & (MeV)          \\
\hline                 
\lam & C+C   & 14$\pm$2 & 0.24 $\pm$0.01$\pm$0.04   & 1.32$\pm$0.05$\pm$0.32 & ---                    & 224$\pm$ 6$\pm$27 & 199$\pm$ 8$\pm$15 \\
     & Si+Si & 37$\pm$3 & 0.88 $\pm$0.04$\pm$0.13   & 3.88$\pm$0.16$\pm$0.56 & ---                    & 267$\pm$16$\pm$32 & 235$\pm$ 9$\pm$16 \\
\hline
\lab & C+C   & 14$\pm$2 & 0.064$\pm$0.003$\pm$0.010 & 0.18$\pm$0.02$\pm$0.03 & 0.92$\pm$0.04$\pm$0.16 & 204$\pm$ 9$\pm$27 & 184$\pm$11$\pm$17 \\
     & Si+Si & 37$\pm$3 & 0.16 $\pm$0.007$\pm$0.038 & 0.49$\pm$0.05$\pm$0.11 & 0.96$\pm$0.13$\pm$0.21 & 230$\pm$10$\pm$30 & 205$\pm$ 9$\pm$17 \\
\hline
\xim & Si+Si & 37$\pm$3 & 0.07 $\pm$0.01 $\pm$0.01  & ---                      & ---                  & 239$\pm$16$\pm$29 & 210$\pm$13$\pm$13 \\
\end{tabular}
\end{ruledtabular}
\end{table*}
%
%
%
%
%



\begin{thebibliography}{99}

\bibitem{RAFELSKI}   J.~Rafelski and B.~M\"{u}ller, 
                     Phys. Rev. Lett. {\bf 48}, 1066 (1982).

\bibitem{NA35LAM}    J.~Bartke et al. (NA35 Collaboration),
                     Z. Phys. C {\bf 48}, 191 (1990).

\bibitem{NA35STR}    T.~Alber et al. (NA35 Collaboration),
                     Z. Phys. C {\bf 64}, 195 (1994).

\bibitem{WA97HYP}    F.~Antinori et al. (WA97 Collaboration),
                     Eur. Phys. J. C {\bf 11}, 79 (1999).

\bibitem{NA57HY158}  F.~Antinori et al. (NA57 Collaboration),
                     J. Phys. G {\bf 32}, 427 (2006).

\bibitem{E802STR}    L.~Ahle et al. (E802 Collaboration),
                     Phys. Rev. C {\bf 60}, 044904 (1999).

\bibitem{E895HYP}    P.~Chung et al. (E895 Collaboration),
                     Phys. Rev. Lett. {\bf 91}, 202301 (2003).

\bibitem{NA49EDHYP}  C.~Alt et al. (NA49 Collaboration),
                     Phys. Rev. C {\bf 78}, 034918 (2008).

\bibitem{NA49CCSISI} C.~Alt et al. (NA49 Collaboration),
                     Phys. Rev. Lett. {\bf 94}, 052301 (2005).

\bibitem{DANOS}      J.~Rafelski and M.~Danos,
                     Phys. Lett. B {\bf 97}, 279 (1980).

\bibitem{TOUNSI}     S.~Hamieh, K.~Redlich, and A.~Tounsi,
                     Phys. Lett. B {\bf 486}, 61 (2000).

\bibitem{CLAUDIA}    C.~H\"{o}hne, F.~P\"{u}hlhofer, and R.~Stock,
                     Phys. Lett. B {\bf 640}, 96 (2006).

\bibitem{WERNER1}    K.~Werner,
                     Phys. Rev. Lett. {\bf 98}, 152301 (2007).

\bibitem{GLAUBER}    R.J.~Glauber,
                     Phys. Rev. {\bf 100}, 242 (1955).

\bibitem{BECATTINI1} F.~Becattini and J.~Manninen,
                     J. Phys. G {\bf 35}, 104013 (2008).

\bibitem{BECATTINI3} F.~Becattini and J.~Manninen,
                     Phys. Lett. B {\bf 673}, 19 (2009).

\bibitem{TIMMINS}    A.R.~Timmins (for the STAR Collaboration),
                     arXiv:0810.0017.

\bibitem{WERNER2}    J.~Aichelin and K.~Werner,
                     arXiv:0810.4465.

\bibitem{NA49NIM}    S.~V.~Afanasiev et al. (NA49 Collaboration), 
                     Nucl. Instrum. Meth. A {\bf 430}, 210 (1999).

\bibitem{PDG}        W.-M.~Yao et al. (Particle Data Group),
                     J. Phys. G {\bf 33}, 1 (2006).

\bibitem{VENUS}      K.~Werner,
                     Phys. Rept. {\bf 232}, 87 (1993).

\bibitem{GEANT3}     Geant---Detector Description and Simulation Tool,  
                     CERN Program Library Long Writeup W5013.

\bibitem{INGRID}     I.~Kraus,
                     PhD thesis, University of Frankfurt (2004).

\bibitem{MICHI}      M.~K.~Mitrovski,
                     PhD thesis, University of Frankfurt (2007).

\bibitem{BECATTINI2} F.~Becattini, J.~Manninen, and M.~Ga\'zdzicki,
                     Phys. Rev. C {\bf 73}, 044905 (2006).

\bibitem{BLASTWAVE}  E.~Schnedermann and U.~Heinz, 
                     Phys. Rev. C {\bf 50}, 1675 (1994).

\bibitem{NA49SDPR1}  T.~Anticic et al. (NA49 Collaboration),
                     Phys. Rev. C {\bf 69}, 024902 (2004).

\bibitem{HSD4}       W.~Cassing,
                     Nucl. Phys. A {\bf 700}, 618 (2002).

\bibitem{CGREINER}   C.~Greiner and S.~Leupold,
                     J. Phys. G {\bf 27}, L95 (2001). 

\bibitem{HSD1}       W.~Ehehalt and W.~Cassing,
                     Nucl. Phys. A {\bf 602}, 449 (1996).

\bibitem{HSD2}       W.~Cassing and E.L.~Bratkovskaya,
                     Phys. Rep. {\bf 308}, 65 (1999).

\bibitem{HSD3}       H.~Weber, E.L.~Bratkovskaya, W.~Cassing, and H.~St\"{o}cker,
                     Phys. Rev. C {\bf 67}, 014904 (2003),
                     and private communication.

\bibitem{URQMD}      M.~Bleicher et al, 
                     J. Phys. G {\bf 25}, 1859 (1999).

\bibitem{URQMD23A}   H.~Petersen, M.~Bleicher, S.~A.~Bass, and H.~St\"{o}cker,
                     arXiv:0805.0567.

\bibitem{URQMD23B}   H.~Petersen, M.~Mitrovski, T.~Schuster, and M.~Bleicher,
                     arXiv:0903.0396.

\bibitem{NA49SDPR2}  C.~Blume et al. (for the NA49 Collaboration),
                     PoS(Confinement08), {\bf 110} (2008),
                     and NA49 publication in preparation.

\bibitem{NA57HY40}   F.~Antinori et al. (NA57 Collaboration),
                     Phys. Lett. B {\bf 595}, 68 (2004).

\bibitem{PPLAM}      F.~Kramer, C.~Strabel, and M.~Ga\'{z}dzicki,
                     arXiv:nucl-ex/0509035.

\bibitem{MAREKPP}    M.~Ga\'{z}dzicki and D.~R\"{o}hrich,
                     Z. Phys. C {\bf 71}, 55 (1996).

\bibitem{TANJAQM}    T.~\v{S}u\v{s}a (for the NA49 Collaboration),
                     Nucl. Phys. A {\bf 698}, 491c (2002).

\end{thebibliography}
\end{document}